\documentclass[]{aa}

\usepackage[switch, modulo]{lineno}

\usepackage{rotating}
\usepackage{graphicx}
\usepackage{sidecap}
\usepackage{subcaption}
\usepackage{url}
\usepackage[colorlinks=true]{hyperref}
\usepackage{soul}
\hypersetup{allcolors=blue}
\usepackage[normalem]{ulem}

\usepackage[kerning=true, tracking=true]{microtype}

\usepackage{lipsum}
\usepackage{txfonts}
\usepackage[]{xcolor}
\usepackage{float}

\bibpunct{(}{)}{;}{a}{}{,}
\DeclareUnicodeCharacter{00A0}{~}
\usepackage[us,12hr]{datetime}

\usepackage{cancel}
\usepackage{color}

\definecolor{deepmagenta}{rgb}{0.8, 0.0, 0.8}
\definecolor{green}{rgb}{0.0, 0.5, 0.1}

\renewcommand{\jobname}{neuralwfa}

\begin{document}

\title{Exploring spectropolarimetric inversions using neural fields}
\subtitle{Solar chromospheric magnetic field under the weak-field approximation}

\author{C. J. D\'iaz Baso
        \inst{1,2}
        \and
        A. Asensio Ramos
        \inst{3,4}
        \and
        J. de la Cruz Rodr\'iguez
        \inst{5}
        \and
        J. M. da Silva Santos
        \inst{6}
        \and\\
        L. Rouppe van der Voort
        \inst{1,2}
        }
        
    \institute{
    Institute of Theoretical Astrophysics,
    University of Oslo, %
    P.O. Box 1029 Blindern, N-0315 Oslo, Norway
    \and
    Rosseland Centre for Solar Physics,
    University of Oslo, %
    P.O. Box 1029 Blindern, N-0315 Oslo, Norway
    \and
    Instituto de Astrof\'isica de Canarias, C/V\'{\i}a L\'actea s/n, E-38205 La Laguna, Tenerife, Spain
    \and 
    Departamento de Astrof\'isica, Universidad de La Laguna, E-38206 La Laguna, Tenerife, Spain
    \and
    Institute for Solar Physics, Dept. of Astronomy, Stockholm University, AlbaNova University Centre, SE-10691 Stockholm, Sweden
    \and
    National Solar Observatory, 3665 Discovery Drive, Boulder, CO 80303, USA
    \\
    \email{carlos.diaz@astro.uio.no}
    }

   \date{Draft: compiled on \today\ at \currenttime~UT}

   \authorrunning{D\'iaz Baso et al.}


\abstract
{
Full-Stokes polarimetric datasets, originating from slit-spectrograph or narrow-band filtergrams, are routinely acquired nowadays. The data rate is increasing with the advent of bi-dimensional spectropolarimeters and observing techniques that allow long-time sequences of high-quality observations. 
There is a clear need to go beyond the traditional pixel-by-pixel strategy in spectropolarimetric inversions by exploiting the spatiotemporal coherence of the inferred physical quantities that contain valuable information about the conditions of the solar atmosphere. 
}
{
We explore the potential of neural networks as a continuous representation of the physical quantities over time and space (also known as neural fields), for spectropolarimetric inversions.}
{
We have implemented and tested a neural field to perform one of the simplest forms of spectropolarimetric inversions, the inference of the magnetic field vector under the weak-field approximation (WFA). By using a neural field to describe the magnetic field vector, we can regularize the solution in the spatial and temporal domain by assuming that the physical quantities are continuous functions of the coordinates. This technique can be trivially generalized to account for more complex inversion methods.
}
{
We have tested the performance of the neural field to describe the magnetic field of a realistic 3D magnetohydrodynamic (MHD) simulation. We have also tested the neural field as a magnetic field inference tool (approach also known as physics-informed neural networks) using the WFA as our radiative transfer model. We investigated the results in synthetic and real observations of the \ion{Ca}{ii} 8542\,\AA\ line. We also explored the impact of other explicit regularizations, such as using the information of an extrapolated magnetic field, or the orientation of the chromospheric fibrils. 
}
{
Compared to the traditional pixel-by-pixel inversion, the neural field approach improves the fidelity of the reconstruction of the magnetic field vector, especially the transverse component. 
This implicit regularization is a way of increasing the effective signal-to-noise of the observations. 
Although it is slower than the pixel-wise WFA estimation, this approach shows a promising potential for depth-stratified inversions, by reducing the number of free parameters and inducing spatio-temporal constraints in the solution.
}

\keywords{Sun: atmosphere -- Sun: chromosphere -- Sun: magnetic fields -- Methods: observational -- Sun: activity -- Radiative transfer}

\maketitle


\section{Introduction}\label{sec:intro}
During the past thirty years, inversion methods have proven to be one of the most robust ways of establishing a quantitative relation between the observed intensities and the underlying physical state of the atmospheric plasma. With the use of fast slit spectropolarimeters and 2D filterpolarimeters, maps of different regions in the solar atmosphere are routinely taken in which the four Stokes parameters are observed at several points along one or several spectral lines. The rate of new high-quality 2D observations is increasing with the advent of bi-dimensional spectropolarimeters and observing techniques that allow long time sequences of high-quality observations \citep{Dominguez-Tagle2022JAI....1150014D,vanNoort2022A&A...668A.149V}. With some exceptions (discussed below), in the overwhelming majority of studies available in the literature, such observations have been interpreted by assuming that all pixels are completely unrelated and by applying the inference techniques (commonly known as inversion codes) on a pixel-by-pixel basis. However, the spatial complexity of the observations is not as chaotic as one would expect from the high dimensionality of the data (spatial, temporal, spectral), but is rather coherent as a consequence of the physical processes that dominate the solar dynamics.

For example, the magnetic field in the chromosphere, where the magnetic pressure is larger than the gas pressure, tends to be rather smooth and slow-varying over space. However, polarimetric signals induced by those chromospheric magnetic fields are particularly weak, and in most cases very close to the detection limit of current instrumentation \citep[e.g.,][]{DiazBaso2019A&A_filament, Yadav2021A&A...649A.106Y}. 
Given these inherent properties, incorporating these characteristics in the inference would significantly help to constrain better the solar atmosphere.

This spectral and spatial coherence has been exploited to effectively reduce the noise in solar and stellar spectropolarimetric observations \citep{MartinezGonzalez2008A&A_pca, DiazBaso2019_denoise} and avoid averaging in time or space, which could lead to a loss of important information. 
The new generation of inversion codes is also starting to make use of this coherency to improve the fidelity of their algorithms. A study by \cite{vanNoort2012A&A} proposed to couple the solution of neighboring pixels using the telescope point spread function. This inspired the recent development of a non-linear spatially-regularized and multi-resolution inversion technique by \citet{delaCruz2019_multires}.

A different approach to take into account spatial coherency was presented by \cite{Asensio2015A&A}, using the concept of sparsity and compressibility, by linearly transforming the physical parameters to a different space in which their representation is compact. They used proximal algorithms \citep{parikh_boyd14} to impose sparsity in the wavelet domain, decreasing the number of free parameters to reproduce the observables while simultaneously favoring spatial coherency. Recent studies (\citealt{delaCruz2019_multires}, \citealt{Morosin2020A&A}, \citealt{delaCruz2024tempRegu}) proposed adding spatio-temporal constraints by explicitly imposing a Tikhonov regularization on the physical parameters, thereby improving the fidelity of the reconstruction. 
At the same time, advances in deep learning have introduced the potential for data-driven regularizations. These methods encode complex priors from simulations into neural networks \citep{AsensioRamos2021A&A...646A...4A, Liaudat2023arXiv231200125L}.
The ideas developed in the context of deep learning have also inspired other works, such as
\citep{Jiri2022A&A...659A.137S, Jiri2024arXiv240720926S}, which propose to solve the 3D inversion problem by including the 3D non-local thermodynamic equilibrium consistency as a regularization, together with additional physical constraints and stochastic
gradient descent techniques to mitigate issues of local minima.
Lastly, automatic differentiation frameworks, such as PyTorch \citep{Paszke2019arXiv}, facilitate the implementation of these ideas by efficiently computing gradients and optimizing models to reproduce observations \citep{DeCeuster2024arXiv240218525D}.

Motivated by the development of new instruments with an increasing field of view, we believe that implicit methods that describe the physical parameters in the whole domain with a compact representation can be of great help to reduce the dimensionality of the problem and, consequently, the computational load of the inference process. Here we investigate a different way of parametrizing the physical quantities by using a neural network as a continuous approximation to introduce spatio-temporal constraints. Recent works have demonstrated the potential of this idea in coronal tomography \citep{AsensioRamos2023SoPh_tomography}, source reconstruction under strong gravitational lenses \citep{Mishra-Sharma2022mla}, magnetic field extrapolations \citep{Jarolim2024ApJ_extrapol2}
and interstellar chemistry \citep{AsensioRamos2024arXiv240602387A}.
These neural networks, usually termed implicit neural representations, neural fields (NF), or coordinate-based representations, are used to map coordinates on the space (or space-time) to coordinate-dependent field quantities. They have many desirable properties: they are efficient in terms of the number of free parameters, they have controllable implicit bias, they produce differentiable quantities that can be part of more elaborate optimizations, and they generate continuous signals that are ideal for imposing spatio-temporal constraints in noisy scenarios. In this work, we will study the case where the magnetic field is inferred under the weak-field approximation \cite[WFA;][]{Landi1973SoPh}, a powerful method to estimate the magnetic field from plage \citep{daSilvaSantos2023ApJ...954L..35D} to flare scenarios \citep{Vissers2021A&A} and simple enough to focus our attention on this particular new implementation. The formalism also remains identical for LTE or non-LTE inversions of chromospheric lines. We plan to extend the use of NFs to more complex radiative transfer models in the near future.

The paper is organized as follows. We start with a brief introduction to their basic principles, and how we implement the new approach to perform spectropolarimetric inversions. Later we show the application of the NF on some examples and introduce some additional explicit regularizations. Finally, we provide a brief discussion about the implications of this work and outline potential extensions and
improvements.

\section{Neural magnetic field reconstruction}

\subsection{Weak-field approximation}

The weak-field approximation \cite[WFA;][]{Landi1973SoPh} is an analytical solution of the radiative transfer equation. This allows us to derive the emerging Stokes $Q$, $U$, and $V$ parameters describing the polarization of the light as a function of Stokes $I$ and its derivatives as a function of wavelength. The fundamental assumptions are that the magnetic field vector is constant with depth and that the splitting induced by the Zeeman effect ($\Delta\lambda_B$) is significantly smaller than the Doppler width of the line ($\Delta\lambda_D$). This weak field regime occurs at different field strengths for different spectral lines (depending on the sensitivity to the magnetic field, the local temperature, etc). 

At first order in the magnetic field strength, the relation between Stokes $V$ and Stokes $I$ 
is given by the following expression:
\begin{equation}
V(\lambda)=-\Delta \lambda_B \, \bar{g} \, \cos{\Theta_B} \, \frac{dI}{d\lambda}
\label{eq:stokesv}
\end{equation}
where $\Delta \lambda _B = 4.6686 \cdot 10^{-13}\lambda_0^2 \, B$, $\Theta_B$ is the inclination of the magnetic field (angle between the observer's line-of-sight and the normal to the solar surface) and $\bar{g}$ is the Land\'e factor \citep{landi_landolfi04}. The central wavelength of the line, $\lambda_0$, is given in \AA\, while the magnetic field strength, $B$, is given in G. The same perturbation analysis allows obtaining Stokes $Q$ and $U$, which only appear at second-order. In particular, we will use the equations that describe the dependence of Stokes $Q$ and $U$ in the wings $(\lambda \gg \Delta\lambda_D)$ of the line:
\begin{eqnarray}\label{eq:wingsQU}
    Q(\lambda_w) &=& \frac{3}{4}\,\Delta\lambda_B^2 \,\bar{G} \,\sin^2{\Theta_B} \,\cos{2\Phi_B} \,\frac{1}{\lambda_w-\lambda_0} \biggl(\frac{dI}{d\lambda}\biggl) \nonumber \\
    U(\lambda_w) &=& \frac{3}{4}\,\Delta\lambda_B^2 \,\bar{G} \,\sin^2{\Theta_B} \,\sin{2\Phi_B} \,\frac{1}{\lambda_w-\lambda_0} \biggl(\frac{dI}{d\lambda}\biggl)
\end{eqnarray}
where $\bar{G}$ is a parameter that gives the magnetic sensitivity of linear polarization to $B_{\perp}$, which depends on the quantum numbers of the transition \citep{landi_landolfi04}. In both equations, there is a term depending on $\Phi_B$, which is the azimuth angle of the magnetic field with respect to a reference direction.


\begin{figure*}[htp!]
\centering
\includegraphics[width=1\linewidth]{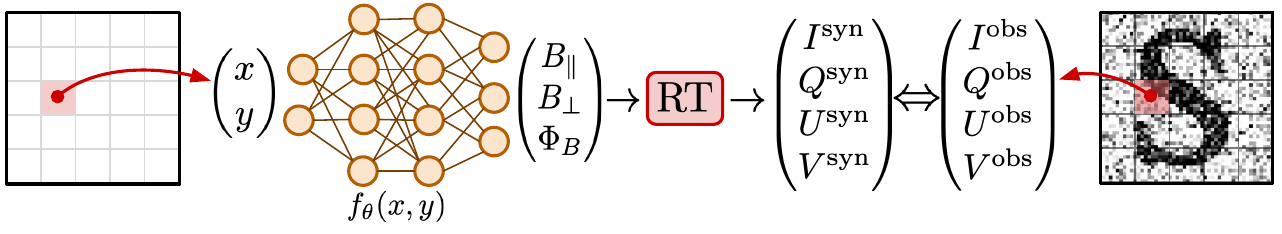}
\caption{Sketch of the NF approach. The physical quantities are described by a neural network that takes the coordinates as input and outputs the physical quantities at that point $(x,y)$. The output of the network is then used to compute the  observables from the model using a radiative transfer (RT) module (WFA in this case). The synthetic observables are compared with the observations and the error is back-propagated through the network to obtain the optimal solution.    
} \label{fig:sketch}
\end{figure*}

\subsection{Magnetic field inference}

Once the model is set, we aim to infer the magnetic field vector from the interpretation of the observations of a set of spectral lines. Assuming that the weak-field approximation can be applied to the observed spectral lines and that observations are corrupted with uncorrelated Gaussian noise, we can use a least-squares estimator (maximum likelihood) to retrieve the magnetic field vector. The merit function $\mathcal{L}$, the well-known $\chi^2$, can be defined for a particular pixel as the mean squared difference of the observed polarization signals and the synthetic ones predicted by the model, normalized by the variance of the noise:
\begin{equation}\label{chi2}
    \mathcal{L} = \mathcal{L}_V + \mathcal{L}_Q + \mathcal{L}_U,
\end{equation}
where
\begin{align}\label{chi2b}
    \mathcal{L}_V &= \sum_i \frac{(V^\mathrm{obs}_i-V^\mathrm{syn}_i)}{\sigma_{V,i}^2}, \nonumber \\
    \mathcal{L}_Q &= \sum_i \frac{(Q^\mathrm{obs}_i-Q^\mathrm{syn}_i)}{\sigma_{Q,i}^2}, \nonumber \\
    \mathcal{L}_U &= \sum_i \frac{(U^\mathrm{obs}_i-U^\mathrm{syn}_i)}{\sigma_{U,i}^2}.
\end{align}

The sub-index $i$ is used as a label for the spectral wavelength points. The previous merit function considers the general case in which the standard deviation is different for Stokes $Q, U$, and $V$. The formal simplicity of Eqs.~(\ref{eq:stokesv}) and (\ref{eq:wingsQU}) is one of the most important reasons why WFA has been useful. This linear dependence on the model parameters allows for an analytical optimization of the $\chi^2$ \citep[see, e.g.,][]{Martinez2012MNRAS}. Since our approach here is more general (taking spatial correlation into account and considering more complex radiative transfer models in the near future), we do not use the analytical optimization of the $\chi^2$. Rather, we consider the numerical optimization of the $\chi^2$ using gradient-based methods.

From a practical point of view, using $B$, $\Theta_B$, and $\Phi_B$ as free variables often leads to problems in the optimization. They are subject to several physical restrictions like the positive definiteness of $B$ or are plagued with discontinuities (the azimuth is periodic in the interval $[0, \pi]$, leading to problems during the optimization). For this reason, we opted to use the following variables, obtained as combinations of $B$, $\Theta_B$, and $\Phi_B$:
\begin{eqnarray}
B_{\parallel} &=& B \cos{\Theta_B} \nonumber \\
B_{Q} &=& (B \sin{\Theta_B})^2 \cos{2 \Phi_B} \nonumber \\
B_{U} &=& (B \sin{\Theta_B})^2 \sin{2 \Phi_B}.
\end{eqnarray}
These three quantities are defined in $(-\infty,+\infty)$ and are continuous functions of the magnetic field vector. They also have the additional advantage of decoupling the problem into three unrelated sub-problems: Stokes $V$ is only dependent on $B_\parallel$, Stokes $Q$ is only dependent on $B_Q$ and Stokes $U$ is only dependent on $B_U$. Therefore, one could solve each sub-problem independently. This is useful, for instance, if one Stokes parameter has much stronger signals than the rest (typically the case of Stokes $V$), which could dominate the optimization process. In that case, one could use different weights for each term in the merit function. Note that this last point is only valid for the WFA and not for the general case.

\subsection{Neural fields}
We propose here a general and powerful technique for the inference of magnetic field (and potentially thermodynamical parameters when using models more complex than the WFA) from observations using neural networks. The general idea is depicted in  Fig.~\ref{fig:sketch} and described in the following. We use a neural network, $f_\theta(x,y)$, to describe the components of the magnetic field vector as a function of the coordinates. In this case, we describe the magnetic field components in a 2D plane in Cartesian coordinates $(x,y)$. These coordinates represent the plane of ``effective'' formation height at which the polarization is generated. In the general case in which stratification and time evolution are taken into account, the input coordinates will be the $(x,y,z,t)$. For convenience, we normalize all coordinates so that they are mapped to the interval $[-1,1]$. The magnetic field components are then given by the following simple, but flexible, fully-connected neural network $f_\theta$:
\begin{equation}
    \vec{B} = (B_\parallel, B_Q, B_U) = f_\theta(\vec{x}),
\end{equation}
with $\theta$ the internal parameters of the neural network and $\vec{x}=(x,y)$.

This approach of describing the magnetic field using NFs has several advantages. First, the number of tunable parameters of the neural network can be potentially fewer than the number of unknowns in all pixels. This might not be especially relevant for the WFA model since there are only 3 unknowns per pixel. However, this will be crucial when applied to stratified inversions where we have many more physical quantities per pixel (temperature, velocity, magnetic field, microturbulence, etc.) in a very dense grid, not only spatial but also in the optical depth domain. Secondly, a NF is a global function in the space. This means that the information introduced by the observation of a single pixel informs the whole solution, leading to a very pronounced regularization effect. This effect is similar to the global character induced by the wavelet decomposition used by \cite{Asensio2015A&A}. Thirdly, a NF is a continuous and differentiable function of the input coordinates. Therefore, the result of the inversion process is a continuous function that can be evaluated at any arbitrary point. Having a continuous and differentiable magnetic field map can be greatly beneficial for the computation of current sheets though spatial derivatives of the magnetic field (see the application to inversion methods of \citealt{2021A&A...656L..20P}), which otherwise will be greatly affected by the (potential) presence of noise and inversion artifacts. Finally, it is straightforward to include any explicit additional regularization term in the optimization that depends on the output or any derivative of the output with respect to the input coordinates.

\begin{figure*}[htp!]
\centering
\includegraphics[width=1\linewidth]{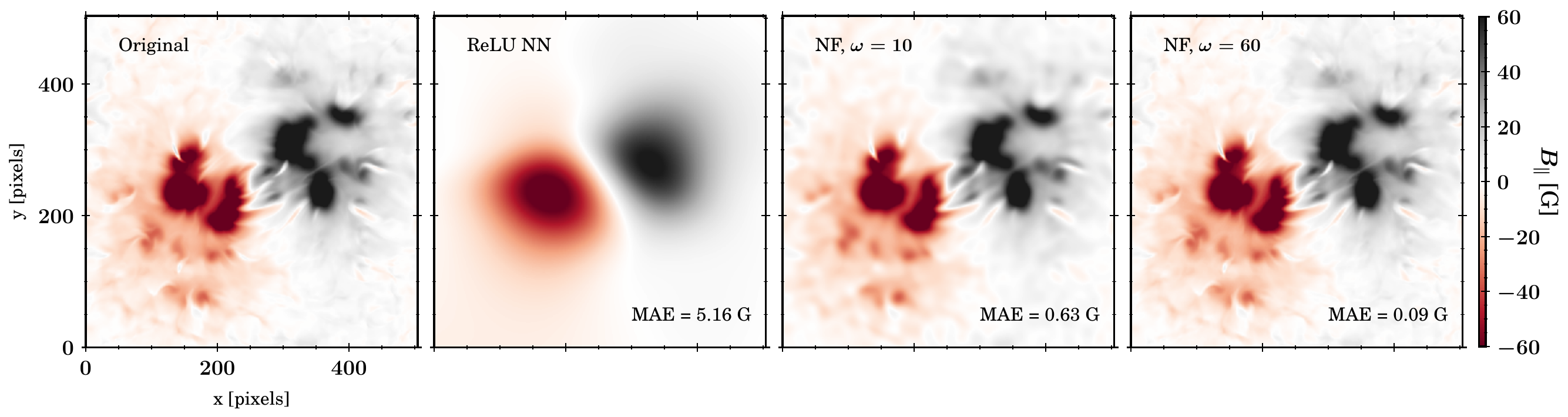}
\caption{Comparison of the parametric representation power of each method. From left to right: the original longitudinal magnetic field at $z=1500$~km in the Bifrost simulation, represented using a ReLU neural network, and using a NF with $\omega=10$ and $\omega=60$.
} \label{fig:representation_power}
\end{figure*}

\subsection{Tests with a 3D rMHD simulation}

\subsubsection{Parametric representation}

Before showing the use of neural networks for the WFA problem, we will test first the performance of the neural network to describe the 2D spatial properties of the physical quantities of interest. To that end, we parameterize the magnetic field of a realistic 3D radiative magneto-hydrodynamics (rMHD) simulation. We have used one snapshot from a publicly available enhanced network simulation \citep{Carlsson2016A&A} performed with the Bifrost code \citep{Gudiksen2011A&A}. Snapshots from this simulation have been extensively used in previous studies to test different diagnostic strategies \citep[e.g.,][]{2012ApJ...749..136L,2013ApJ...772...90L,2017A&A...597A..46S, Jurcak2018A&A}. The longitudinal magnetic field at  1500\,km from the mean continuum formation layer is shown in the left panel of Fig.~\ref{fig:representation_power}. This panel shows a chromospheric landscape with elongated magnetic features connecting two opposite-polarity patches. This image can be represented with the aid of a NF by minimizing the following loss:
\begin{equation}
    \mathcal{L}_{B_\parallel} = \sum_{x,y} \left( f_\theta(x,y)-B_\parallel(x,y) \right)^2,
    \label{eq:loss_bparallel}
\end{equation}
where we utilize a fully-connected NN with ReLU \citep[Rectified Linear Unit;][]{relu69} activation functions. 
Note that we do not \textit{train} the neural network in the traditional manner for generalization across multiple datasets; instead, each network is uniquely optimized for a specific dataset to function as a tailored parametric tool, not as a generalized predictive model.
The converged solution is found in the second panel of Fig.~\ref{fig:representation_power}. The solution only approximates the general behavior of the original magnetic field distribution but is too smooth. This smoothness is a direct consequence of the implicit bias of NN, also known as spectral bias \citep{Rahaman2018arXiv}, which prevents standard networks from learning high-frequency functions\footnote{Surprisingly, this implicit bias is one of the key reasons for the success of deep learning, in which NNs focus on the general properties of the data manifolds, instead of focusing on the details.}. This is an active area of research, and several strategies have been proposed to alleviate this problem and introduce an implicit bias towards high-frequency signals. One of the most successful strategies is to pass the input coordinates through a Fourier feature mapping $\gamma(\vec{x})$, which allows the NF to correctly generate high spatial frequencies \citep{Tancik2020arXiv}. This mapping projects the input coordinates onto a high-dimensional space with a set of trigonometric functions:
\begin{equation}
    \gamma(\vec{x}) = \left[\cos(2 \pi \vec{G} \vec{x}), \sin(2 \pi \vec{G} \vec{x})\right]^T
\end{equation}
where $\vec{G} \sim \mathcal{N}(0, \omega)$, i.e., each entry of the vector $\vec{G}$ is a frequency sampled from a Gaussian distribution with zero mean and standard deviation $\omega$. The parameter $\omega$ controls the range of spatial frequencies that the network can reproduce. This will be helpful as a regularization, as we show later. After computing the Fourier features, we pass them through the neural network to obtain the magnetic field. By using a Fourier mapping, we can efficiently reproduce both low and high spatial frequencies, being able to represent the magnetic field with a high fidelity. The third and fourth panels of Fig.~\ref{fig:representation_power} show the results of using the Fourier features with different values of $\omega$. Using $\omega=10$ (third panel), allows us to capture the lower spatial frequencies, but by increasing it to $\omega=60$ (fourth panel) we can reliably reproduce all the high-frequency spatial details of the magnetic field map. In the bottom right of each panel, we also quantify the mean absolute error (MAE) between the original simulation and the different results, decreasing from an average error of 5\,G with the ReLU NN to 0.09\,G using the NF ($\omega=60$). We note that other strategies to alleviate the spectral bias exist. It is worth mentioning that the use of suitable activation functions, such as sinusoids, has also been shown to be very effective in generating high-frequency functions \citep[SIREN;][]{Sitzmann2020arXiv}. All these strategies can be seen as a nonlinear extension of Fourier series.

It is important to stress the fact that the number of learnable parameters of the NF is not dependent on the number of pixels in the observations, but rather on the properties of the spatial distribution of the physical quantities. In the example shown in Fig.~\ref{fig:representation_power}, although the number of pixels is $500 \times 500$, the number of parameters of the network is ten times smaller, amounting to just 25k. This compact representation is arguably associated with the fact that the magnetic field in the chromosphere, where the magnetic pressure is larger than the gas pressure, tends to be rather smooth and slow-varying over space. This is not to be expected for other quantities, such as the temperature or the velocity, which vary at much smaller spatial scales.

\begin{figure}[t]
\centering
\includegraphics[width=1\linewidth]{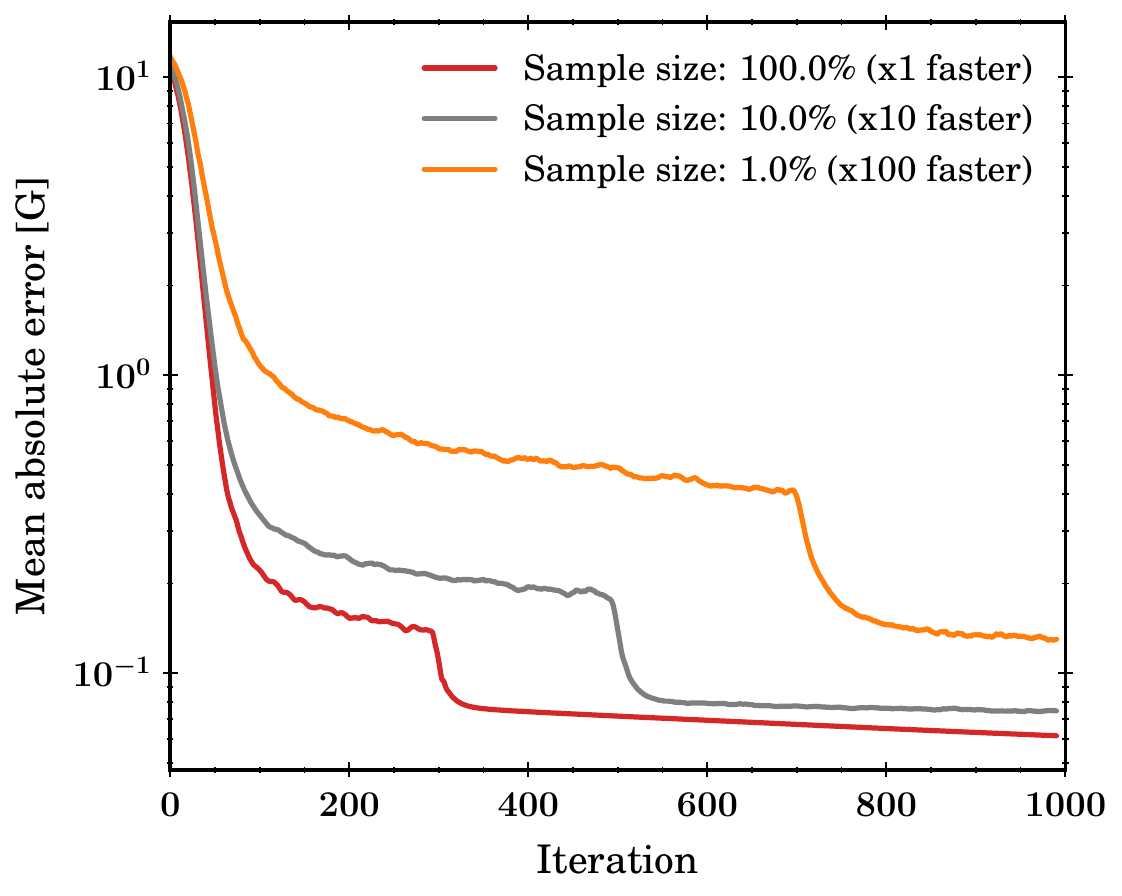}
\caption{Performance of the NF when  evaluating different batches of pixels randomly chosen in the field of view in every iteration.} \label{fig:sample_size}
\end{figure}

The representations shown in Fig.~\ref{fig:representation_power} have been trained by optimizing the loss function of Eq. (\ref{eq:loss_bparallel}) computed for all pixels in the field-of-view (FoV). However, since the NF is a global function in $\vec{x}$, the loss in a given pixel contains some information about the properties of the NF in the surroundings. For this reason, one can train the NF using mini-batches of pixels randomly chosen in the FoV, instead of summing over the whole FoV. To show this, Fig.~\ref{fig:sample_size} shows the mean absolute error as a function of iteration when different mini-batches are selected for each iteration. We can see that the convergence properties when using 10\% of the total pixels are already as good as those obtained when using the whole FoV, but 10 times faster in terms of computing time. The sudden decrease in the merit function during the optimization is produced by the scheduler, a module that decreases the learning rate (step size) if there is no improvement after some iterations.

\begin{figure*}[htp!]
\centering
\includegraphics[width=1\linewidth]{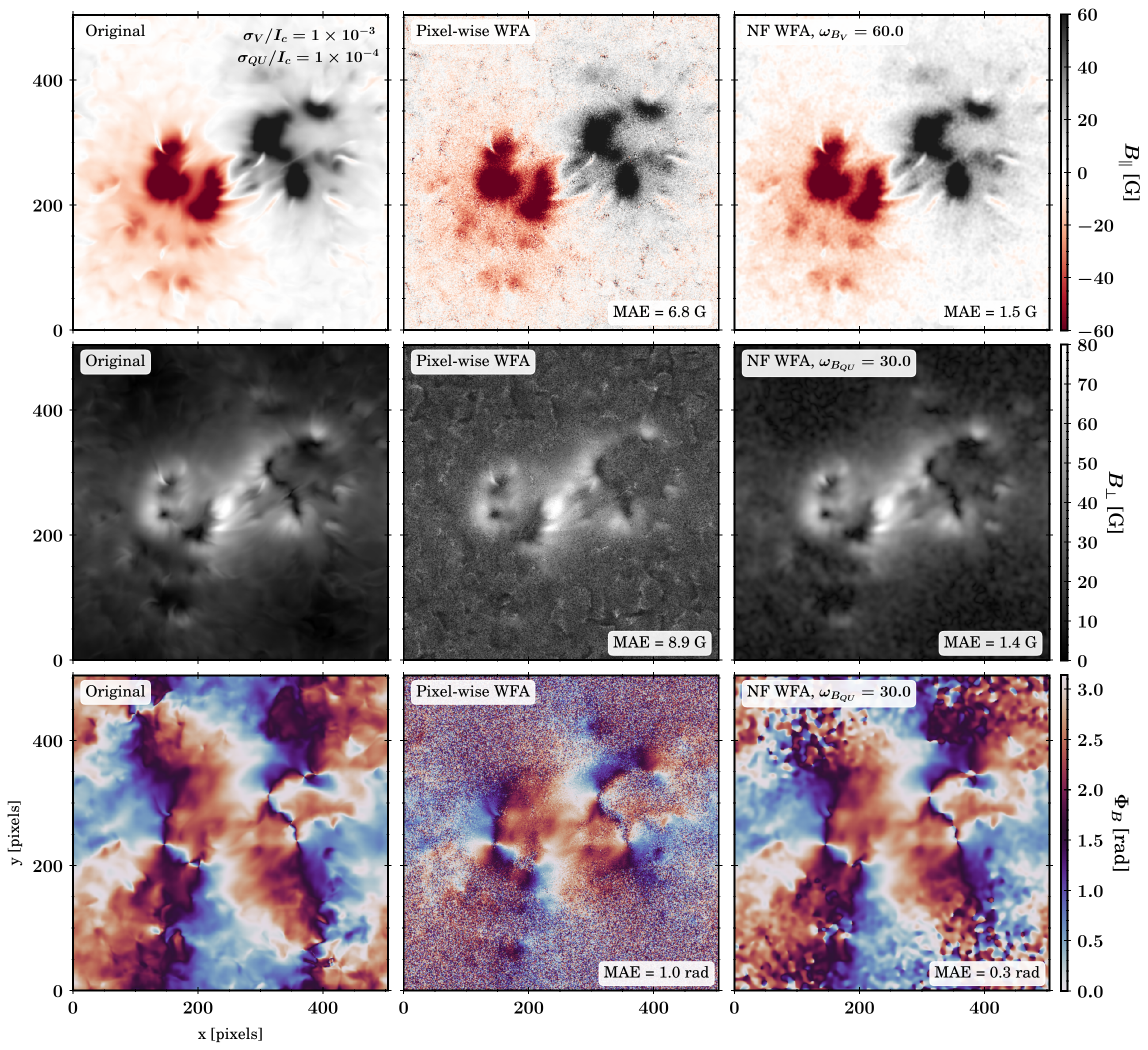}
\caption{Comparison of the reconstruction of the magnetic field vector from the synthetic \ion{Ca}{ii} 8542\AA\ spectra calculated from the simulation (in rows: line-of-sight component, transverse component, and azimuth angle). From left to right: original magnetic field from the simulation, magnetic field inferred by the pixel-wise WFA, and results using the WFA NF.} \label{fig:simu_mosaic}
\end{figure*}

\subsubsection{Magnetic field reconstruction}

The previous experiment demonstrates that a NF can correctly recover the details of the quantities of interest by their direct observation. But in spectropolarimetry, we only have access to the Stokes profiles and one has to pass through the radiative transfer model to infer the physical quantities. Here we demonstrate that this can be done by interpreting synthetic observations of the \ion{Ca}{ii} 8542\,\AA\ line. We have chosen this line because the WFA approximation on this line is reliable for field strengths up to $\sim$1200\,G \citep{Centeno2018ApJ...866...89C}. For this test, we have created synthetic observables from the Bifrost simulation and we run our WFA NF to reproduce the polarization signals that emerge from the simulation. Our goal is to assess the improvement delivered by our new method when compared to the traditional pixel-by-pixel WFA. To this end, and to discard any influence of the line formation details, we follow \cite{Morosin2020A&A} and set the magnetic field in each pixel equal to its vertical average from $z=1000$ km to $z=1500$ km. Heights are measured assuming that $z=0$ km corresponds to the mean continuum formation layer. We compute the synthetic observables using the non-local thermodynamic equilibrium (NLTE) radiative transfer STiC code\footnote{\url{https://github.com/jaimedelacruz/stic}} \citep{delaCruz2016,delaCruz2019_STiC} and add uncorrelated Gaussian noise to the Stokes $Q$, $U$, and $V$ profiles. It is important to note that the average magnetic field of this particular snapshot is about 50--100~G, a value much lower than the typical magnetic field found in plage \citep[$\sim$400 G, ][]{Pietrow2020A&A,Morosin2022A&A...664A...8M}. Consequently, to avoid the signal being buried in the noise, we use a standard deviation of the noise lower than the typical one found in solar observations.

The architecture chosen for the NF is a residual network \citep[ResNet,][]{ResNet2015} with 2 residual blocks, 64 nodes per layer, and the ELU activation function \citep[Exponential Linear Unit;][]{2015arXiv151107289C}. This architecture has proven its effectiveness by showing efficient training and good performance in capturing complex patterns. For the Fourier feature mapping, we have used $\omega=60$ for the longitudinal magnetic field and $\omega=30$ for the transverse components. The reason is that the latter is typically more affected by the noise. Using a smaller value of $\omega$ encourages the NF not to fit the noise, which is of very high spatial frequency. The magnetic field generated by the NF is used to compute the synthetic observables via the WFA approximation, which are then compared with the observations. In particular, for the longitudinal magnetic field, and using Eq. (\ref{eq:stokesv}), we have used the following loss function:
\begin{eqnarray}
\mathcal{L}_V = \sum_{x,y} \sum_\lambda \left[\left(- C_V B_\parallel(x,y|\theta)  \frac{dI^\mathrm{obs}_{x,y}}{d\lambda}\right)- V^\mathrm{obs}_{x,y}\right]^2
\label{eq:lossv}
\end{eqnarray}
where $C_{V}=4.6686 \times 10^{-13} \bar{g} \lambda_0^2$ is a constant for the line of interest, $B_\parallel(x,y|\theta)$ is the NF used to represent the longitudinal field, $dI^\mathrm{obs}_{x,y}/d\lambda$ is the derivative of the observed Stokes $I$ profile at pixel $(x,y)$, and $V^\mathrm{obs}_{x,y}$ is the observed Stokes $V$ signal at the same pixel. The derivative of the intensity does not change during the optimization, so it can be pre-calculated and stored before the optimization process occurs.

For the transverse components, using Eq. (\ref{eq:wingsQU}), the loss functions are given by: 
\begin{eqnarray}
\mathcal{L}_Q = \sum_{x,y} \sum_\lambda \left[\left( C_{QU}\cdot B_Q(x,y |\theta) \cdot \frac{1}{\lambda-\lambda_0}\frac{dI^\mathrm{obs}_{x,y}}{d\lambda}\right)- Q^\mathrm{obs}_{x,y}\right]^2
\label{eq:lossq}
\end{eqnarray}
\begin{eqnarray}
\mathcal{L}_U = \sum_{x,y} \sum_\lambda \left[\left( C_{QU}\cdot B_U(x,y |\theta) \cdot \frac{1}{\lambda-\lambda_0}\frac{dI^\mathrm{obs}_{x,y}}{d\lambda}\right)- U^\mathrm{obs}_{x,y}\right]^2
\label{eq:lossu}
\end{eqnarray}
where $C_{QU}=1.6347 \times 10^{-25} \lambda_0^4 \bar{G}$. Thanks to the definition of the $B_Q$ and $B_U$ variables, the losses can be optimized separately. This helps in imposing different implicit or explicit regularization methods for every variable.

The optimization of every loss function with respect to $\theta$ is carried out using the Adam optimizer with a learning rate of $3\cdot10^{-4}$ for 100$-$500 epochs (depending on the dataset). The derivatives for the backpropagation through the network and the physical model (WFA) are carried out with the automatic differentiation package PyTorch. Finally, after convergence of the NFs, one can recover the full magnetic field vector from $B_\parallel$, $B_Q$, and $B_U$ by a final transformation:
\begin{eqnarray}
B_\parallel(x,y) &=& B_\parallel(x,y|\theta) \nonumber \\
B_{\perp}(x,y) &=& \left[B_Q(x,y|\theta)^2 + B_U(x,y|\theta)^2 \right]^{1/4} \nonumber \\
\Phi_B(x,y) &=& \arctan\left(\frac{B_U(x,y|\theta)}{B_Q(x,y|\theta)}\right).
\end{eqnarray}


Figure \ref{fig:simu_mosaic} shows the reconstruction of the magnetic field vector in the synthetic case used before in Fig.~\ref{fig:representation_power}. The first column shows the original magnetic field vector, given in terms of $B_\parallel$, $B_\perp$ , and $\Phi_B$. The second column shows the inferred magnetic field using the traditional pixel-based WFA and the third row shows the inferred magnetic field using the NFs. The quality of the results is affected, fundamentally, by the noise level and the regularization properties of NF. We have tested the performance of the NF for different noise levels and values of $\omega$, finding the same consistent behavior. The results summarized in Fig.~\ref{fig:simu_mosaic} are for the particular configuration with a noise level of $10^{-3}$ for Stokes $V$ and $10^{-4}$ for Stokes $Q$ and $U$, both given in units of the continuum intensity. The longitudinal field inferred with the NF is very similar to that of the traditional WFA in general terms. The NF strongly damps the high-frequency components of the magnetic field associated with the presence of noise, but the overall structure is very similar. The key reason is that the longitudinal magnetic field is a quantity that is well constrained by the observations even in the presence of noise because the Stokes $V$ signals are typically above the noise and the estimated value is statistically unbiased and coincides with the original value \citep[see][]{Martinez2012MNRAS}.

The NF produces a much better result for the transverse component and the azimuth of the magnetic field, which are much closer to the real ones than those obtained with the traditional WFA. In this case, the implicit spatial regularization of the NF is able to provide a much better solution. The pixel-by-pixel WFA tends to overestimate the transversal component of the magnetic field, producing a background component over the FoV which is proportional to the noise level. This effect is well-known and produced because the maximum-likelihood solution is biased \cite{Martinez2012MNRAS}. The standard approach to deal with this bias is to avoid the regions where this effect has a strong impact on any subsequent analysis. Here we show that the NF is able to provide a better estimate of the magnetic field by finding a solution that is more coherent with the surroundings. This produces a mitigation of the bias of the transversal component. At the bottom right of each panel, we have quantified both approaches, retrieving an average error 2.5$-$4.5 times lower with the NF WFA compared with the pixel-wise approach. Our results confirm the advantages of the spatial regularization found by \cite{Morosin2020A&A}. 
Note that these small errors do not fully encapsulate the complexities found in actual observations, which are influenced by various systematic factors such as noise or spectral and spatial degradation. Nonetheless, within our idealized setting, the inference using the NF WFA is indicative of the potential improvements.
Lastly, our calculations provide two more conclusions. The first one is that, if we decrease the value of $\omega$, the NF tends to produce an excessively smooth solution. Although this is bad in general, it can be an advantage in very noisy observations. The second one is the finding that the larger the noise level, the better the reconstruction of the magnetic field using the NF WFA is compared to the traditional WFA (see Fig.~\ref{fig:simu_mosaic_appendix} in the Appendix for a comparison using an increased noise level).

\begin{figure*}[htp!]
\centering
\includegraphics[width=\linewidth]{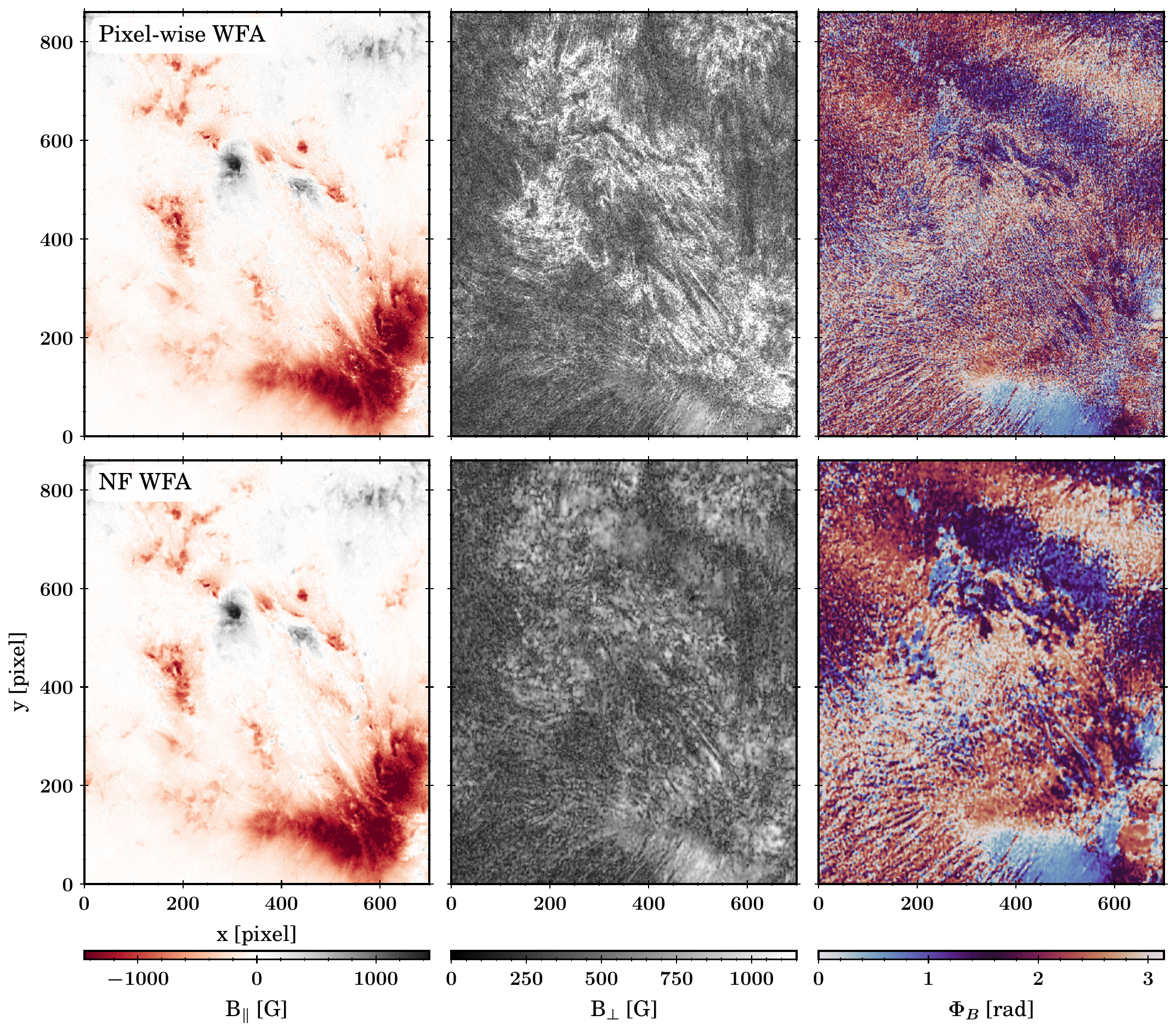}
\caption{
Magnetic field reconstruction from a \ion{Ca}{ii} 8542\AA\ observation using the pixel-wise WFA (top row) and using the NF WFA with $\omega=120$ for $B_\parallel$ and $\omega=80$ for $B_\perp$ \& $\Phi_B$ (bottom row). Columns show the magnetic field in terms of the longitudinal component, the transverse component, and the azimuth angle.}
\label{fig:figureJ}
\end{figure*}

\section{Application to observations}
\subsection{Example with real observations}\label{sec:goodcase}

After confirming with simulations that the WFA NF is able to recover the magnetic field with a higher fidelity than the standard pixel-by-pixel WFA in noisy cases, we apply it to real observations. We have used observations \citep{Leenaarts2018A&A, Yadav2023ApJ_ERF} from the active region NOAA 12593 observed on 2016-09-19 between 09:31:29 UT and 09:57:03 UT with the CRISP \citep{Scharmer2008} instrument at the Swedish 1-m Solar Telescope \citep[SST;][]{Scharmer2003}. The selection of the regularization frequencies $\omega$ was guided by empirical testing: we experimented with different values to find an optimal balance that minimizes noise while preserving critical features of the magnetic field, particularly the transverse component. The results from the analysis of the first time frame are shown in Fig.~\ref{fig:figureJ}. The first row shows the results with the pixel-wise WFA and the second row displays those obtained when applying the WFA NF. As expected, since the Stokes $V$ signals are well above the noise, the inference with both methods is very similar. In contrast, the pixel-wise WFA tends to estimate the transverse magnetic field larger than 1000~G in many locations of the field of view. This is the mentioned bias effect, with very strong values because in the presence of a hot magnetic canopy, the line source function is very shallow in the chromosphere and the resulting line profile has a flat line core (see Appendix A of \citealt{Morosin2020A&A}). In other words, the pixel-wise WFA compensates the almost-zero Stokes $I$ derivative with a very large magnetic field value. This effect is more noticeable in Fabry-P\'erot observations due to the poor sampling of Stokes $I$, as shown in \cite{DiazBaso2023A&A...673A..35D}. The NF, on the other hand, is able to provide a more coherent solution in the spatial domain without using unrealistic values. The maximum values are now below 800~G, where most of them are in the edge of the lower-right polarity (penumbra of the sunspot) and in between the small polarities in the FoV.

\subsection{Challenging case and additional spatial regularization}

\begin{figure*}[htp!]
\centering
\includegraphics[width=\linewidth]{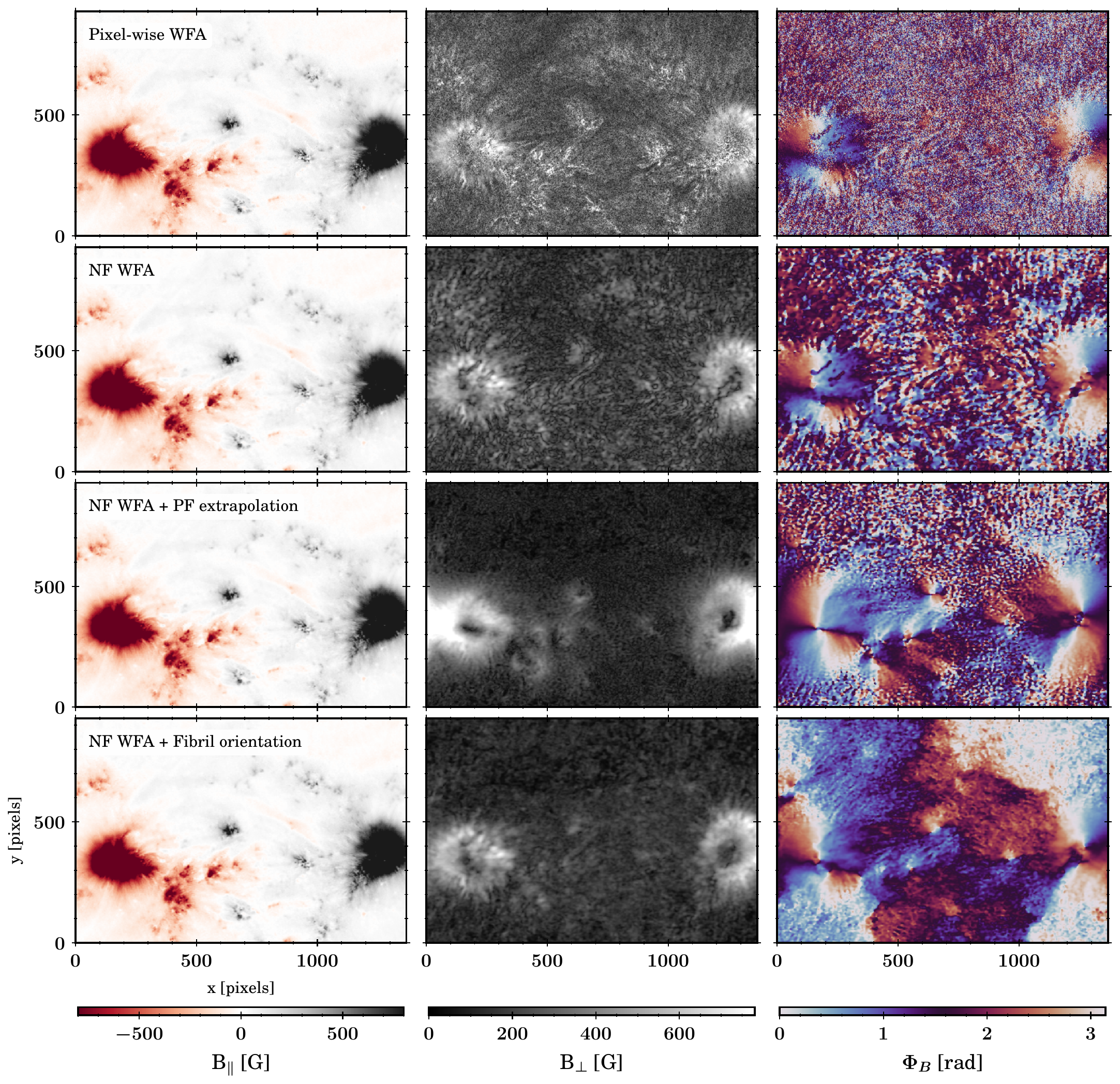}
\caption{
Magnetic field inference from a \ion{Ca}{ii} 8542\AA\ observation using the pixel-wise WFA, the NF WFA, and two additional approaches introducing an explicit regularization term: using the information of a potential extrapolation of the magnetic field (third row) and using the orientation of the fibrils (fourth row). The first column shows the longitudinal magnetic field, the second column shows the transverse magnetic field and the third column shows the azimuth of the magnetic field.
} \label{fig:mosaic_regularizations}
\end{figure*}

We have used observations of the active region NOAA AR 12723 recorded on 2018-09-30 in the \ion{Ca}{ii} 8542\,\AA\ line by \cite{Vissers2022A&A_wfavsmhm} with the CRISP instrument at the SST consisting of a mosaic of four overlapping pointings. This target was selected to explore the behavior of the NF when the magnetic field is particularly weak, but with a simple enough connectivity to investigate what additional information we could add to the inference problem. In this case, the NF WFA does a job similar to the pixel-wise WFA because the magnetic field is mainly concentrated at the sunspots and decreases rapidly for increasing distances from the sunspots. However, the formulation of the inversion process in terms of a NF allows us to include additional regularizations.

The results of the inversions are shown in Fig.~\ref{fig:mosaic_regularizations}. The longitudinal magnetic field, the transverse magnetic field, and the azimuth of the magnetic field are shown in the first, second, and third columns, respectively. The results clearly show that a NF is marginally regularizing the inference. The quality of the inferred magnetic field far from the sunspots is similar between both the NF WFA and the pixel-wise WFA. The estimation of the magnetic field is slightly more spatially coherent, as expected, but the overall structure is very similar. This is a clear example of a limiting case in which there is not enough information to constrain the solution.

In order to improve the magnetic field estimation, we have implemented an explicit regularization. It is implemented by guiding the NF to simultaneously obey another physical constraint, apart from fitting the spectropolarimetric observations. This constraint is implemented as a regularization term in the loss function parameterized with a hyperparameter $\lambda$, which penalizes the distance between the magnetic field configuration and the magnetic field configuration from the external source of information. Given that the longitudinal magnetic field is well constrained, we will  focus here on adding a regularization to the transverse component, i.e.:
\begin{eqnarray}
\mathcal{L} = \mathcal{L}_Q + \mathcal{L}_U + \lambda \mathcal{L}_{\rm reg}.
\end{eqnarray}
As a first example, the third row of Fig.~\ref{fig:mosaic_regularizations} shows the result when $\mathcal{L}_{\rm reg}=\mathcal{L}_{\rm pot}$, where $\mathcal{L}_{\rm pot}$ is the mean squared difference between the NF and a pre-calculated potential field extrapolation $B_\mathrm{pot}$ from the well-constrained longitudinal magnetic field component. One would want to calculate the difference between these two magnetic fields defining $\mathcal{L}_{\rm pot}$ following:
\begin{eqnarray}
\mathcal{L}_\mathrm{pot} = \sum_{x,y} \left( B_\perp - B_{\perp,\rm pot} \right)^2+ \sum_{x,y} \left( \Phi_B - \Phi_{B,\rm pot} \right)^2.
\end{eqnarray}

However, as we have not disambiguated the magnetic field azimuth, we have to calculate the distance between the corresponding analog quantities $B_{U,\rm pot}$, and $B_{Q,\rm pot}$ :
\begin{eqnarray}
\mathcal{L}_\mathrm{pot} = \sum_{x,y} \left( B_Q - B_{Q,\rm pot} \right)^2+ \sum_{x,y} \left( B_U - B_{U,\rm pot} \right)^2.
\end{eqnarray}

When the value of $\lambda$ is increased during the optimization, the NF starts to incorporate the information of the potential field extrapolation. However, even with a small regularization value, we can already see from Fig.~\ref{fig:mosaic_regularizations} that the azimuth of the magnetic field becomes more spatially coherent, but the new transverse magnetic field becomes much stronger than the one estimated only from reproducing the polarization signals. This experiment shows that this particular region is far from being potential and incorporating this information will worsen the fits as soon as we increase the value of $\lambda$. More complex extrapolations can be performed but the idea here is to show how to incorporate external information into the inference problem.

Finally, we also explore the idea of incorporating the orientation of the chromospheric fibrils as if they were aligned with the magnetic field. In fact, this idea has been used to improve nonlinear force-free modeling of coronal fields \citep{Wiegelmann2008SoPh..247..249W}, given the limitations of the force-free assumption of the photospheric boundary \citep{DeRosa2015ApJ...811..107D}. Extrapolations performed starting from a chromospheric vector boundary condition \citep{Fleishman2017ApJ...839...30F} or starting from the photosphere and adding even an incomplete set of chromospheric magnetic field data \citep{Fleishman2019ApJ...870..101F} can measurably improve the reconstruction of the coronal magnetic field, connectivity, and electric currents. This information can potentially also improve the inference of the magnetic field, especially in areas away from strong magnetic field concentrations.

To calculate the distance between the magnetic field azimuth $\Phi_B$ and the direction of the fibrils $\Phi_{\rm fib}$, we need to convert all angles to corresponding points on the unit circle to avoid the ambiguity problem, and then we can compute the distance of these points. The fourth row of Fig.~\ref{fig:mosaic_regularizations} shows the result when $\mathcal{L}_{\rm reg}=\mathcal{L}_{\rm fib}$, with 
\begin{eqnarray}
\mathcal{L}_\mathrm{fib} = \sum_{x,y} \left[ \sin(2\Phi_B) - \sin(2\Phi_{\rm fib}) \right]^2 + \nonumber \\ \sum_{x,y} \left[ \cos(2\Phi_B) - \cos(2\Phi_{\rm fib}) \right]^2,
\end{eqnarray}
i.e., the mean squared difference between the NF azimuth and the direction of the fibrils, as calculated from the intensity image. In order to infer the orientation of the fibrils from the observations we have used the following procedure: {\sc i}) we have used the core of the \ion{Ca}{ii} 8542\,\AA\ line to detect the fibrils in the chromosphere\footnote{We note that some pre-processing is usually performed to enhance the skeleton of the fibrils before the Sobel filters. However, we did not find a significant difference, so we left out this extra step.}, 
{\sc ii}) we apply a Sobel operator in each axis and take the arctangent to retrieve the orientation of the fibrils, which is collapsed to the range (0,180) degrees to avoid the 180 degrees ambiguity and {\sc iii}) a Gaussian filter is applied to remove small artifacts at the edges of the fibrils (see Fig.~\ref{fig:orientation} in the Appendix for more information).
As a result of the inference, Fig.~\ref{fig:mosaic_regularizations} shows a magnetic field that is aligned in general with the fibrils while still reproducing the polarization signals. The middle panel shows that after incorporating the orientation of the fibrils, the transverse component remains almost the same. In fact, the estimation of the orientation of the fibrils fails in the umbra where fibrils are not visible (see Fig.~\ref{fig:orientation}) but the strong polarization signals are enough to compensate for that. This guided inferred magnetic field can be a much better boundary condition for coronal field extrapolations. 


Other potential regularizations that we should explore in the future are forcing the divergence-free condition or the suppression of strong electric currents. Both constraints can be computed from derivatives of the output of the neural network with respect to the input coordinates, which can be computed efficiently with techniques from automatic differentiation. This could allow us to resolve the Zeeman-180 degree azimuth ambiguity at the same time we are reproducing the spectra.

\begin{figure*}[htp!]
\centering
\includegraphics[width=1\linewidth]{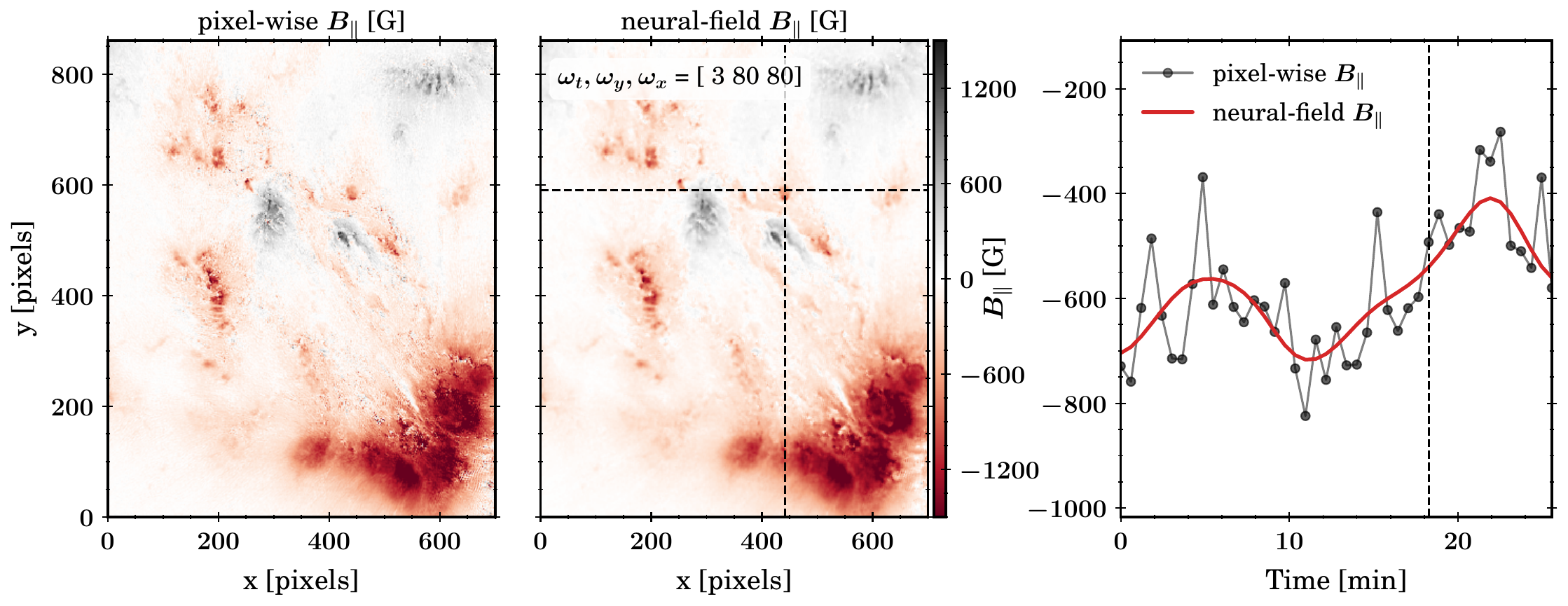}
\includegraphics[width=1\linewidth]{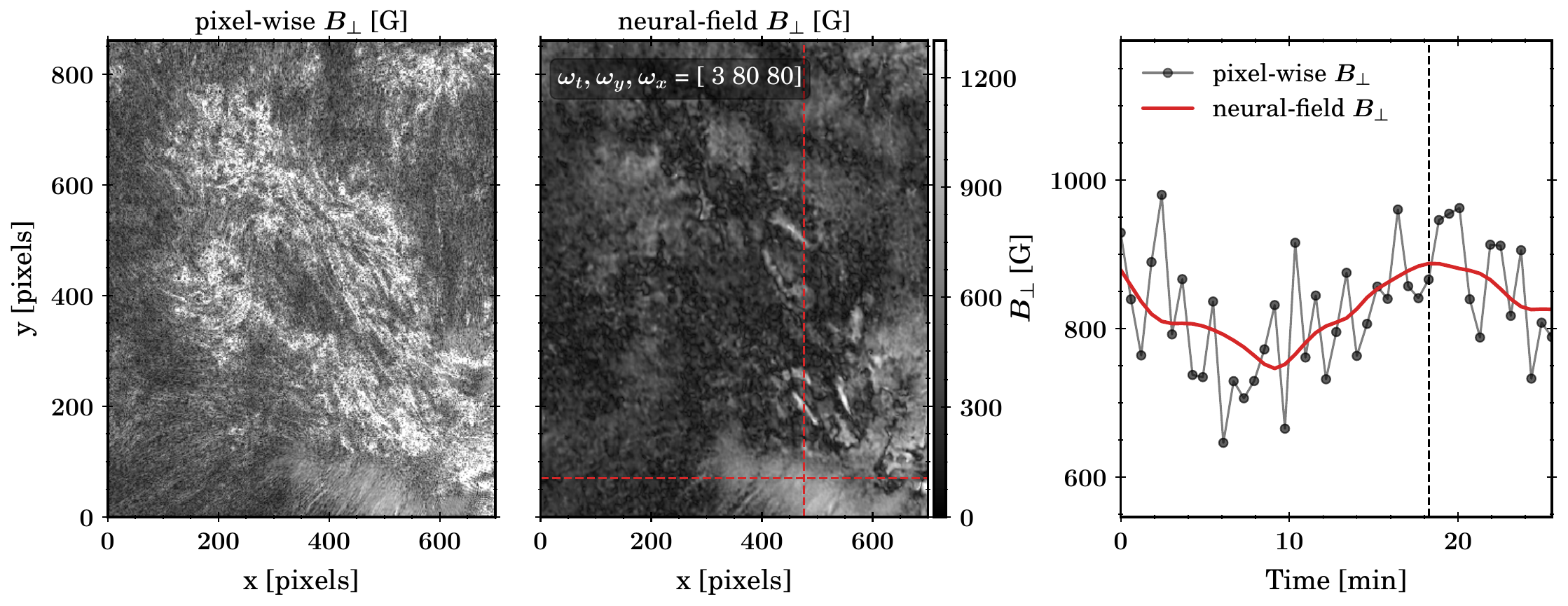}
\caption{
Longitudinal magnetic field (upper row) and transverse component (lower row) inferred using the pixel-wise WFA and the temporal WFA NF for a time series of the active region NOAA 12593 observed on 2016-09-19 with the SST/CRISP. The rightmost column shows the temporal evolution of the magnetic field for a particular pixel indicated as the intersection of the dashed lines in the middle panel.
} \label{fig:temporal}
\end{figure*}

\subsection{Temporal regularization}

The NF can be easily extended to the case where the magnetic field is not static but evolves with time. By imposing temporal coherence in the solution, one can obtain a better estimate of the magnetic field, as shown by \cite{delaCruz2024tempRegu}. From a technical point of view, adding the time dependence can be done by adding $t$ as an additional input parameter of the NF:
\begin{equation}
    \vec{B} = (B_\parallel, B_Q, B_U) = f_\theta(\vec{x},t).
\end{equation}
Apart from this change, inferring the magnetic field proceeds exactly as before, with the only change that the optimization process requires minimizing the loss function over the spatio-temporal domain. 
The number of unknowns in the explicit temporal regularization of \cite{delaCruz2024tempRegu} scales linearly with the number of observed time steps ($n_t$) since it infers the value of the magnetic field at all $n_x n_y$ pixels of the field-of-view. This requires solving a very large linear system of equations of size $n_x n_y n_t \times n_x n_y n_t$. On the contrary, the NF produces a much more compact representation of the magnetic field because it only requires adding a few weights from the input layer to the Fourier feature mapping layer. Since the magnetic field contains higher frequencies in the spatial directions than in the temporal direction, we implement the Fourier feature mapping with $\omega_t$ for the time and $\omega_{xy}$ for the spatial coordinates, with $\omega_t < \omega_{xy}$.

To showcase this approach, we have used a 26-minute time series of the active region NOAA 12593 described in Sec.~\ref{sec:goodcase}. We optimize the temporal WFA NF on the \ion{Ca}{ii} 8542\,\AA\ observations of this time series. For this particular example, we have used a temporal regularization of $\omega_t=3$ and a spatial regularization of $\omega_{xy}=80$. Higher values prevented the neural network from converging correctly. The results of the inference are shown in Fig.~\ref{fig:temporal} for the longitudinal magnetic field and the transverse component. The rightmost column shows the temporal evolution of the magnetic field component for the particular pixel shown in the middle panel. 

In the case of the longitudinal magnetic field (first row of Fig.~\ref{fig:temporal}), the NF is able to capture the details of the spatial distribution of the emerging flux region. In the extracted pixel, the NF is also able to capture the temporal evolution of the magnetic field. The fluctuations of pixel-wise inversion are of the order of $\sim$100~G, which are compatible with the noise amplitude of the observations. This makes us confident that the implicit bias introduced in the NF with the small value of $\omega_t$ correctly captures the time variation of the longitudinal component of the magnetic field. However, since this particular region is very dynamic, one could argue that a less restrictive temporal regularization could be needed to properly capture the details of the small-scale changes and a higher $\omega_t$ would be necessary in this type of scenario.

In the case of the transverse magnetic field (second row of Fig.~\ref{fig:temporal}), there is a much larger difference between the output of the NF and that of the traditional pixel-wise WFA. The traditional WFA tends to estimate a transverse magnetic field larger than 1000~G in the center of the FoV far from the sunspot, where we have weak polarization signals. This is again the bias produced by the noise described before. The NF, on the other hand, is able to provide a more coherent solution both spatially and temporally. Both approaches retrieve very similar magnetic field values where the signals are stronger. When compared with the previous section where only spatial regularization was employed, temporal regularization is particularly effective in reducing the background noise \citep{delaCruz2024tempRegu}.


\section{Summary and conclusions}\label{sec:conclusions}

In this study, we investigated the use of neural fields to parameterize the magnetic field for magnetic field inference. We have shown the capabilities of NFs to solve static and time-dependent magnetic field inferences. This implementation\footnote{Our implementation is publicly available in the following repository:  \url{https://github.com/cdiazbas/neural_wfa}.} comes with several important advantages. 

First, by reformulating the problem as a global problem (instead of pixel-wise independent problems), the NF can produce a much more compact representation, decreasing the number of parameters to optimize. This reduction generally depends on both the spatial complexity of the inferred physical quantities and the efficiency of the chosen neural architecture to represent such complexity. We believe that, for three-dimensional cases like the inversion of Stokes profiles with stratified atmospheres, NFs will represent an exceptionally compact representation of the physical conditions \citep[see, for example,][]{AsensioRamos2023SoPh_tomography}. So, we anticipate that NFs will become crucial as new modern instruments and telescopes are providing bigger FoV of more complex data (e.g. at the SST, the recently installed CRISP cameras offer a FoV diameter of 87\arcsec\ and the forthcoming CRISP2 will offer a FoV of 2 arcminutes).

Second, using a NF to describe the magnetic field as a continuous differentiable function allows us to obtain a better estimation of the magnetic field in places where the signals are buried in the noise, without smearing out the details of locations where the signals are strong. Choosing adequate regularization frequencies $\omega$ is crucial to obtain results that are not overly smooth. These frequencies should be chosen so that the quality of the fit is not degraded but still maintains sharp gradients in locations where the signals significantly impact the quality of the fit. In fact, NFs seem to be key in the inference of the transverse magnetic field, which the pixel-wise WFA tends to overestimate. This new implementation can help us to better understand and quantify chromospheric heating and its relation with the strength of the horizontal magnetic field in the low chromosphere \citep{Leenaarts2018A&A,DiazBaso2021_heating,daSilvaSantos2022A&A}.

Third, the present work is based on a simple forward model for the Stokes profiles. A NF version of the WFA is not realistically competitive in terms of speed with other WFA implementations. 
For instance, the explicitly regularized WFA implementation of \citet{Morosin2020A&A} can estimate the magnetic field of the FoV in less than a second, while the NF version requires $\sim$30~seconds. The temporal regularization takes also more time, almost $\sim$3 minutes in an off-the-shelf GPU, compared with $\sim$10~minutes in a CPU. 
The main reason for this large processing time is the relatively slow convergence of the NF given that it is optimized using first-order gradient-based techniques. On the other hand, the loss function in spectropolarimetric inversions is often optimized using (quasi-) second-order methods such as the Levenberg-Marquardt algorithm. Second-order methods require the construction of a very large approximate Hessian matrix that can impact memory consumption. For instance, for an observation with $1000 \times 1000$ pixels and 50 time steps and using a Milne Eddington model ($\sim$10 free parameters), the Hessian matrix is of size $5\times10^8 \times 5\times10^8$. In any case, the global Hessian matrix is relatively sparse. Depending on the sparsity degree of the Hessian matrix, the resolution of the coupled linear system of equations can pose a severe computational challenge, even when efficient iterative methods are used (e.g., GMRES or BICGSTAB). Imposing nearest neighbor regularization yields a very compact and sparse matrix that can be efficiently inverted. More dense cases, originating for example from the inclusion of an extended spatial point-spread-function or horizontal coupling by 3D radiative transfer effects, could be trivially included in this framework, whereas it is not trivial to model the inverse problem.

Fourth, modern automatic differentiation frameworks, like PyTorch, allow us to seamlessly use more complex forward models, such as a Milne-Eddington model, or solve the non-LTE inversion problem in a stratified atmosphere. In \citealt{2019A&A...626A.102A}, we already anticipated that traditional \textit{predictive} neural networks, despite their remarkable speed, are not explicitly fitting Stokes profiles and the introduction of a differentiable forward synthesis would notably increase the optimization time. Considering that the forward models are the bottleneck in complex spectropolarimetric inversions, emulators of the radiative transfer model \citep[e.g.,][]{AsensioRamos2024arXiv240602387A} emerges as a crucial strategy to speed up the calculations. Additionally, these frameworks allow us to seamlessly speed up the calculations using GPUs. Adding extra regularization terms to guide the solution is straightforward in these frameworks. As mentioned, by improving the estimation of the magnetic field at the chromospheric level, we can perform better coronal magnetic field extrapolations, and the chromospheric fibrils contain valuable information about the non-potentiality of the magnetic field which should be integrated into the inference \citep{Jing2011ApJ...739...67J}. Apart from the ones used in this work (alignment of the magnetic field with the chromospheric fibrils or similarity to a pre-computed magnetic field extrapolation), one can think of adding Tikhonov regularization on the physical quantities or their spatial derivatives, sparsity constraints using linear transformation like wavelets \citep[e.g.,][]{Asensio2015A&A} or a divergence-free constrain. By extending this model to include additional constraints we could potentially resolve the Zeeman-180 degree azimuth ambiguity directly within the inversion process \citep{Jiri2022A&A...659A.137S,Jiri2024arXiv240720926S}. This would enhance the accuracy and reliability of the inferred magnetic field vectors.

Finally, a natural extension would be a probabilistic reconstruction (i.e., providing the uncertainty of the magnetic field) by using gradient-based variational inference, as shown by \cite{Mishra-Sharma2022mla} in the context of gravitational lens reconstruction. This could be accomplished, for example, by modeling the magnetic field distribution using a normalizing flow \citep{DiazBaso2022A&A...659A.165D}. Another point to explore is the accuracy in reproducing very high-frequency features in the observations because our Fourier Feature layer becomes less stable when we increase the complexity of the problem. A way of mitigating this problem can be using a multi-scale representation \citep{Dolean2023arXiv230605486D, Saragadam2022arXiv220203532S} or add higher frequencies progressively as the training progresses \citep{AsensioRamos2024arXiv240602387A}. Having demonstrated the potential of the method in a proof-of-principle setting, we leave these extensions to future work.

In summary, NFs allow us to improve the reconstruction of the magnetic field properties of the solar atmosphere, especially suitable for imaging spectropolarimeters with large fields of view and a scarce wavelength sampling but also for integral-field spectropolarimeters \citep{vanNoort2022A&A...668A.149V,Rouppe2023A&A...673A..11R} where, although the FoV is smaller, the temporal cadence is very high and the temporal regularization will hold much better. These properties make NFs valuable for the data taken with the next generation of telescopes such as the existing Daniel K. Inouye Solar Telescope \citep[DKIST;][]{Rimmele2020SoPh_dkist} and the upcoming European Solar Telescope \citep[EST;][]{2022A&A...666A..21Q}.

\begin{acknowledgements}

%
CJDB acknowledges M. L. DeRosa for valuable discussions on future application of NF inversion methods.
This research is supported by the Research Council of Norway, project number 325491, 
and through its Centres of Excellence scheme, project number 262622. 
AAR acknowledges financial support from the Agencia Estatal de Investigaci\'on del Ministerio de Ciencia, Innovaci\'on
y Universidades (MCIU/AEI) and the European Regional Development Fund (ERDF) through project PID2022-136563NB-I00.
This project has been funded by the European Union through the European Research Council (ERC) under the Horizon Europe program (MAGHEAT, grant agreement 101088184). 
The NSO is operated by the Association of Universities for
Research in Astronomy, Inc., under cooperative agreement
with the National Science Foundation.
The Swedish 1-m Solar Telescope is operated on the island of La Palma by the Institute for Solar Physics of Stockholm University in the Spanish Observatorio del Roque de los Muchachos of the Instituto de Astrof\'isica de Canarias. The Institute for Solar Physics is supported by a grant for research infrastructures of national importance from the Swedish Research Council (registration number 2021-00169). 
We acknowledge the community effort devoted to the development of the following open-source packages that were used in this work: numpy (\url{numpy.org}), matplotlib (\url{matplotlib.org}), scipy (\url{scipy.org}), astropy (\url{astropy.org}) and sunpy (\url{sunpy.org}).
This research has made use of NASA's Astrophysics Data System Bibliographic Services.
\end{acknowledgements}

\bibliographystyle{aa}
\bibliography{references}

\begin{thebibliography}{65}
\expandafter\ifx\csname natexlab\endcsname\relax\def\natexlab#1{#1}\fi

\bibitem[{{Asensio Ramos}(2023)}]{AsensioRamos2023SoPh_tomography}
{Asensio Ramos}, A. 2023, \href{http://dx.doi.org/10.1007/s11207-023-02226-2}{\color{blue}\solphys}, \href{https://ui.adsabs.harvard.edu/abs/2023SoPh..298..135A}{298, 135}

\bibitem[{{Asensio Ramos} \& {de la Cruz Rodr{\'\i}guez}(2015)}]{Asensio2015A&A}
{Asensio Ramos}, A. \& {de la Cruz Rodr{\'\i}guez}, J. 2015, \href{http://dx.doi.org/10.1051/0004-6361/201425508}{\color{blue}\aap}, \href{https://ui.adsabs.harvard.edu/abs/2015A&A...577A.140A}{577, A140}

\bibitem[{{Asensio Ramos} \& {D{\'\i}az Baso}(2019)}]{2019A&A...626A.102A}
{Asensio Ramos}, A. \& {D{\'\i}az Baso}, C.~J. 2019, \href{http://dx.doi.org/10.1051/0004-6361/201935628}{\color{blue}\aap}, \href{https://ui.adsabs.harvard.edu/abs/2019A&A...626A.102A}{626, A102}

\bibitem[{{Asensio Ramos} \& {Pall{\'e}}(2021)}]{AsensioRamos2021A&A...646A...4A}
{Asensio Ramos}, A. \& {Pall{\'e}}, E. 2021, \href{http://dx.doi.org/10.1051/0004-6361/202040066}{\color{blue}\aap}, \href{https://ui.adsabs.harvard.edu/abs/2021A&A...646A...4A}{646, A4}

\bibitem[{{Asensio Ramos} {et~al.}(2024){Asensio Ramos}, {Westendorp Plaza}, {Navarro-Almaida}, {Rivi{\`e}re-Marichalar}, {Wakelam}, \& {Fuente}}]{AsensioRamos2024arXiv240602387A}
{Asensio Ramos}, A., {Westendorp Plaza}, C., {Navarro-Almaida}, D., {et~al.} 2024, \href{http://dx.doi.org/10.1093/mnras/stae1432}{\color{blue}\mnras}, \href{https://ui.adsabs.harvard.edu/abs/2024MNRAS.531.4930A}{531, 4930}

\bibitem[{{Carlsson} {et~al.}(2016){Carlsson}, {Hansteen}, {Gudiksen}, {Leenaarts}, \& {De Pontieu}}]{Carlsson2016A&A}
{Carlsson}, M., {Hansteen}, V.~H., {Gudiksen}, B.~V., {Leenaarts}, J., \& {De Pontieu}, B. 2016, \href{http://dx.doi.org/10.1051/0004-6361/201527226}{\color{blue}\aap}, \href{https://ui.adsabs.harvard.edu/abs/2016A&A...585A...4C}{585, A4}

\bibitem[{{Centeno}(2018)}]{Centeno2018ApJ...866...89C}
{Centeno}, R. 2018, \href{http://dx.doi.org/10.3847/1538-4357/aae087}{\color{blue}\apj}, \href{https://ui.adsabs.harvard.edu/abs/2018ApJ...866...89C}{866, 89}

\bibitem[{{Clevert} {et~al.}(2015){Clevert}, {Unterthiner}, \& {Hochreiter}}]{2015arXiv151107289C}
{Clevert}, D.-A., {Unterthiner}, T., \& {Hochreiter}, S. 2015, \href{https://ui.adsabs.harvard.edu/abs/2015arXiv151107289C}{\href{http://dx.doi.org/10.48550/arXiv.1511.07289}{\color{blue}arXiv e-prints}, arXiv:1511.07289}

\bibitem[{{da Silva Santos} {et~al.}(2022){da Silva Santos}, {Danilovic}, {Leenaarts}, {de la Cruz Rodr{\'\i}guez}, {Zhu}, {White}, {Vissers}, \& {Rempel}}]{daSilvaSantos2022A&A}
{da Silva Santos}, J.~M., {Danilovic}, S., {Leenaarts}, J., {et~al.} 2022, \href{http://dx.doi.org/10.1051/0004-6361/202243191}{\color{blue}\aap}, \href{https://ui.adsabs.harvard.edu/abs/2022A&A...661A..59D}{661, A59}

\bibitem[{{da Silva Santos} {et~al.}(2023){da Silva Santos}, {Reardon}, {Cauzzi}, {Schad}, {Mart{\'\i}nez Pillet}, {Tritschler}, {W{\"o}ger}, {Hofmann}, {Stauffer}, \& {Uitenbroek}}]{daSilvaSantos2023ApJ...954L..35D}
{da Silva Santos}, J.~M., {Reardon}, K., {Cauzzi}, G., {et~al.} 2023, \href{http://dx.doi.org/10.3847/2041-8213/acf21f}{\color{blue}\apjl}, \href{https://ui.adsabs.harvard.edu/abs/2023ApJ...954L..35D}{954, L35}

\bibitem[{{De Ceuster} {et~al.}(2024){De Ceuster}, {Ceulemans}, {Decin}, {Danilovich}, \& {Yates}}]{DeCeuster2024arXiv240218525D}
{De Ceuster}, F., {Ceulemans}, T., {Decin}, L., {Danilovich}, T., \& {Yates}, J. 2024, \href{https://ui.adsabs.harvard.edu/abs/2024arXiv240218525D}{\href{http://dx.doi.org/10.48550/arXiv.2402.18525}{\color{blue}arXiv e-prints}, arXiv:2402.18525}

\bibitem[{{de la Cruz Rodr{\'\i}guez}(2019)}]{delaCruz2019_multires}
{de la Cruz Rodr{\'\i}guez}, J. 2019, \href{http://dx.doi.org/10.1051/0004-6361/201936635}{\color{blue}\aap}, \href{https://ui.adsabs.harvard.edu/abs/2019A&A...631A.153D}{631, A153}

\bibitem[{{de la Cruz Rodr{\'\i}guez} \& {Leenaarts}(2024)}]{delaCruz2024tempRegu}
{de la Cruz Rodr{\'\i}guez}, J. \& {Leenaarts}, J. 2024, \href{http://dx.doi.org/10.1051/0004-6361/202348810}{\color{blue}\aap}, \href{https://ui.adsabs.harvard.edu/abs/2024A&A...685A..85D}{685, A85}

\bibitem[{{de la Cruz Rodr{\'\i}guez} {et~al.}(2016){de la Cruz Rodr{\'\i}guez}, {Leenaarts}, \& {Asensio Ramos}}]{delaCruz2016}
{de la Cruz Rodr{\'\i}guez}, J., {Leenaarts}, J., \& {Asensio Ramos}, A. 2016, \href{http://dx.doi.org/10.3847/2041-8205/830/2/L30}{\color{blue}\apjl}, \href{https://ui.adsabs.harvard.edu/abs/2016ApJ...830L..30D}{830, L30}

\bibitem[{{de la Cruz Rodr{\'\i}guez} {et~al.}(2019){de la Cruz Rodr{\'\i}guez}, {Leenaarts}, {Danilovic}, \& {Uitenbroek}}]{delaCruz2019_STiC}
{de la Cruz Rodr{\'\i}guez}, J., {Leenaarts}, J., {Danilovic}, S., \& {Uitenbroek}, H. 2019, \href{http://dx.doi.org/10.1051/0004-6361/201834464}{\color{blue}\aap}, \href{https://ui.adsabs.harvard.edu/abs/2019A&A...623A..74D}{623, A74}

\bibitem[{{DeRosa} {et~al.}(2015){DeRosa}, {Wheatland}, {Leka}, {Barnes}, {Amari}, {Canou}, {Gilchrist}, {Thalmann}, {Valori}, {Wiegelmann}, {Schrijver}, {Malanushenko}, {Sun}, \& {R{\'e}gnier}}]{DeRosa2015ApJ...811..107D}
{DeRosa}, M.~L., {Wheatland}, M.~S., {Leka}, K.~D., {et~al.} 2015, \href{http://dx.doi.org/10.1088/0004-637X/811/2/107}{\color{blue}\apj}, \href{https://ui.adsabs.harvard.edu/abs/2015ApJ...811..107D}{811, 107}

\bibitem[{{D{\'\i}az Baso} {et~al.}(2022){D{\'\i}az Baso}, {Asensio Ramos}, \& {de la Cruz Rodr{\'\i}guez}}]{DiazBaso2022A&A...659A.165D}
{D{\'\i}az Baso}, C.~J., {Asensio Ramos}, A., \& {de la Cruz Rodr{\'\i}guez}, J. 2022, \href{http://dx.doi.org/10.1051/0004-6361/202142018}{\color{blue}\aap}, \href{https://ui.adsabs.harvard.edu/abs/2022A&A...659A.165D}{659, A165}

\bibitem[{{D{\'\i}az Baso} {et~al.}(2019{\natexlab{a}}){D{\'\i}az Baso}, {de la Cruz Rodr{\'\i}guez}, \& {Danilovic}}]{DiazBaso2019_denoise}
{D{\'\i}az Baso}, C.~J., {de la Cruz Rodr{\'\i}guez}, J., \& {Danilovic}, S. 2019{\natexlab{a}}, \href{http://dx.doi.org/10.1051/0004-6361/201936069}{\color{blue}\aap}, \href{https://ui.adsabs.harvard.edu/abs/2019A&A...629A..99D}{629, A99}

\bibitem[{{D{\'\i}az Baso} {et~al.}(2021){D{\'\i}az Baso}, {de la Cruz Rodr{\'\i}guez}, \& {Leenaarts}}]{DiazBaso2021_heating}
{D{\'\i}az Baso}, C.~J., {de la Cruz Rodr{\'\i}guez}, J., \& {Leenaarts}, J. 2021, \href{http://dx.doi.org/10.1051/0004-6361/202040111}{\color{blue}\aap}, \href{https://ui.adsabs.harvard.edu/abs/2021A&A...647A.188D}{647, A188}

\bibitem[{{D{\'\i}az Baso} {et~al.}(2019{\natexlab{b}}){D{\'\i}az Baso}, {Mart{\'\i}nez Gonz{\'a}lez}, \& {Asensio Ramos}}]{DiazBaso2019A&A_filament}
{D{\'\i}az Baso}, C.~J., {Mart{\'\i}nez Gonz{\'a}lez}, M.~J., \& {Asensio Ramos}, A. 2019{\natexlab{b}}, \href{http://dx.doi.org/10.1051/0004-6361/201834790}{\color{blue}\aap}, \href{https://ui.adsabs.harvard.edu/abs/2019A&A...625A.128D}{625, A128}

\bibitem[{{D{\'\i}az Baso} {et~al.}(2023){D{\'\i}az Baso}, {Rouppe van der Voort}, {de la Cruz Rodr{\'\i}guez}, \& {Leenaarts}}]{DiazBaso2023A&A...673A..35D}
{D{\'\i}az Baso}, C.~J., {Rouppe van der Voort}, L., {de la Cruz Rodr{\'\i}guez}, J., \& {Leenaarts}, J. 2023, \href{http://dx.doi.org/10.1051/0004-6361/202346230}{\color{blue}\aap}, \href{https://ui.adsabs.harvard.edu/abs/2023A&A...673A..35D}{673, A35}

\bibitem[{{Dolean} {et~al.}(2023){Dolean}, {Heinlein}, {Mishra}, \& {Moseley}}]{Dolean2023arXiv230605486D}
{Dolean}, V., {Heinlein}, A., {Mishra}, S., \& {Moseley}, B. 2023, \href{https://ui.adsabs.harvard.edu/abs/2023arXiv230605486D}{\href{http://dx.doi.org/10.48550/arXiv.2306.05486}{\color{blue}arXiv e-prints}, arXiv:2306.05486}

\bibitem[{{Dominguez-Tagle} {et~al.}(2022){Dominguez-Tagle}, {Collados}, {Lopez}, {Cedillo}, {Esteves}, {Grassin}, {Vega}, {Mato}, {Quintero}, {Rodriguez}, {Regalado}, \& {Gonzalez}}]{Dominguez-Tagle2022JAI....1150014D}
{Dominguez-Tagle}, C., {Collados}, M., {Lopez}, R., {et~al.} 2022, \href{http://dx.doi.org/10.1142/S2251171722500143}{\color{blue}Journal of Astronomical Instrumentation}, \href{https://ui.adsabs.harvard.edu/abs/2022JAI....1150014D}{11, 2250014}

\bibitem[{{Fleishman} {et~al.}(2019){Fleishman}, {Mysh'yakov}, {Stupishin}, {Loukitcheva}, \& {Anfinogentov}}]{Fleishman2019ApJ...870..101F}
{Fleishman}, G., {Mysh'yakov}, I., {Stupishin}, A., {Loukitcheva}, M., \& {Anfinogentov}, S. 2019, \href{http://dx.doi.org/10.3847/1538-4357/aaf384}{\color{blue}\apj}, \href{https://ui.adsabs.harvard.edu/abs/2019ApJ...870..101F}{870, 101}

\bibitem[{{Fleishman} {et~al.}(2017){Fleishman}, {Anfinogentov}, {Loukitcheva}, {Mysh'yakov}, \& {Stupishin}}]{Fleishman2017ApJ...839...30F}
{Fleishman}, G.~D., {Anfinogentov}, S., {Loukitcheva}, M., {Mysh'yakov}, I., \& {Stupishin}, A. 2017, \href{http://dx.doi.org/10.3847/1538-4357/aa6840}{\color{blue}\apj}, \href{https://ui.adsabs.harvard.edu/abs/2017ApJ...839...30F}{839, 30}

\bibitem[{Fukushima(1969)}]{relu69}
Fukushima, K. 1969, \href{http://dx.doi.org/10.1109/TSSC.1969.300225}{\color{blue}IEEE Transactions on Systems Science and Cybernetics}, 5, 5

\bibitem[{{Gudiksen} {et~al.}(2011){Gudiksen}, {Carlsson}, {Hansteen}, {Hayek}, {Leenaarts}, \& {Mart{\'\i}nez-Sykora}}]{Gudiksen2011A&A}
{Gudiksen}, B.~V., {Carlsson}, M., {Hansteen}, V.~H., {et~al.} 2011, \href{http://dx.doi.org/10.1051/0004-6361/201116520}{\color{blue}\aap}, \href{https://ui.adsabs.harvard.edu/abs/2011A&A...531A.154G}{531, A154}

\bibitem[{{He} {et~al.}(2015){He}, {Zhang}, {Ren}, \& {Sun}}]{ResNet2015}
{He}, K., {Zhang}, X., {Ren}, S., \& {Sun}, J. 2015, \href{https://ui.adsabs.harvard.edu/abs/2015arXiv151203385H}{arXiv e-prints, arXiv:1512.03385}

\bibitem[{{Jarolim} {et~al.}(2024){Jarolim}, {Tremblay}, {Rempel}, {Molnar}, {Veronig}, {Thalmann}, \& {Podladchikova}}]{Jarolim2024ApJ_extrapol2}
{Jarolim}, R., {Tremblay}, B., {Rempel}, M., {et~al.} 2024, \href{http://dx.doi.org/10.3847/2041-8213/ad2450}{\color{blue}\apjl}, \href{https://ui.adsabs.harvard.edu/abs/2024ApJ...963L..21J}{963, L21}

\bibitem[{{Jing} {et~al.}(2011){Jing}, {Yuan}, {Reardon}, {Wiegelmann}, {Xu}, \& {Wang}}]{Jing2011ApJ...739...67J}
{Jing}, J., {Yuan}, Y., {Reardon}, K., {et~al.} 2011, \href{http://dx.doi.org/10.1088/0004-637X/739/2/67}{\color{blue}\apj}, \href{https://ui.adsabs.harvard.edu/abs/2011ApJ...739...67J}{739, 67}

\bibitem[{{Jur{\v{c}}{\'a}k} {et~al.}(2018){Jur{\v{c}}{\'a}k}, {{\v{S}}t{\v{e}}p{\'a}n}, {Trujillo Bueno}, \& {Bianda}}]{Jurcak2018A&A}
{Jur{\v{c}}{\'a}k}, J., {{\v{S}}t{\v{e}}p{\'a}n}, J., {Trujillo Bueno}, J., \& {Bianda}, M. 2018, \href{http://dx.doi.org/10.1051/0004-6361/201732265}{\color{blue}\aap}, \href{https://ui.adsabs.harvard.edu/abs/2018A&A...619A..60J}{619, A60}

\bibitem[{{Landi Degl'Innocenti} \& {Landi Degl'Innocenti}(1973)}]{Landi1973SoPh}
{Landi Degl'Innocenti}, E. \& {Landi Degl'Innocenti}, M. 1973, \href{http://dx.doi.org/10.1007/BF00152807}{\color{blue}\solphys}, \href{https://ui.adsabs.harvard.edu/abs/1973SoPh...31..299L}{31, 299}

\bibitem[{{Landi Degl'Innocenti} \& {Landolfi}(2004)}]{landi_landolfi04}
{Landi Degl'Innocenti}, E. \& {Landolfi}, M. 2004, {Polarization in Spectral Lines} (Kluwer Academic Publishers)

\bibitem[{{Leenaarts} {et~al.}(2012){Leenaarts}, {Carlsson}, \& {Rouppe van der Voort}}]{2012ApJ...749..136L}
{Leenaarts}, J., {Carlsson}, M., \& {Rouppe van der Voort}, L. 2012, \href{http://dx.doi.org/10.1088/0004-637X/749/2/136}{\color{blue}\apj}, \href{https://ui.adsabs.harvard.edu/abs/2012ApJ...749..136L}{749, 136}

\bibitem[{{Leenaarts} {et~al.}(2018){Leenaarts}, {de la Cruz Rodr{\'\i}guez}, {Danilovic}, {Scharmer}, \& {Carlsson}}]{Leenaarts2018A&A}
{Leenaarts}, J., {de la Cruz Rodr{\'\i}guez}, J., {Danilovic}, S., {Scharmer}, G., \& {Carlsson}, M. 2018, \href{http://dx.doi.org/10.1051/0004-6361/201732027}{\color{blue}\aap}, \href{https://ui.adsabs.harvard.edu/abs/2018A&A...612A..28L}{612, A28}

\bibitem[{{Leenaarts} {et~al.}(2013){Leenaarts}, {Pereira}, {Carlsson}, {Uitenbroek}, \& {De Pontieu}}]{2013ApJ...772...90L}
{Leenaarts}, J., {Pereira}, T.~M.~D., {Carlsson}, M., {Uitenbroek}, H., \& {De Pontieu}, B. 2013, \href{http://dx.doi.org/10.1088/0004-637X/772/2/90}{\color{blue}\apj}, \href{https://ui.adsabs.harvard.edu/abs/2013ApJ...772...90L}{772, 90}

\bibitem[{{Liaudat} {et~al.}(2023){Liaudat}, {Mars}, {Price}, {Pereyra}, {Betcke}, \& {McEwen}}]{Liaudat2023arXiv231200125L}
{Liaudat}, T.~I., {Mars}, M., {Price}, M.~A., {et~al.} 2023, \href{https://ui.adsabs.harvard.edu/abs/2023arXiv231200125L}{\href{http://dx.doi.org/10.48550/arXiv.2312.00125}{\color{blue}arXiv e-prints}, arXiv:2312.00125}

\bibitem[{{Mart{\'\i}nez Gonz{\'a}lez} {et~al.}(2008){Mart{\'\i}nez Gonz{\'a}lez}, {Asensio Ramos}, {Carroll}, {Kopf}, {Ram{\'\i}rez V{\'e}lez}, \& {Semel}}]{MartinezGonzalez2008A&A_pca}
{Mart{\'\i}nez Gonz{\'a}lez}, M.~J., {Asensio Ramos}, A., {Carroll}, T.~A., {et~al.} 2008, \href{http://dx.doi.org/10.1051/0004-6361:200809719}{\color{blue}\aap}, \href{https://ui.adsabs.harvard.edu/abs/2008A&A...486..637M}{486, 637}

\bibitem[{{Mart{\'\i}nez Gonz{\'a}lez} {et~al.}(2012){Mart{\'\i}nez Gonz{\'a}lez}, {Manso Sainz}, {Asensio Ramos}, \& {Belluzzi}}]{Martinez2012MNRAS}
{Mart{\'\i}nez Gonz{\'a}lez}, M.~J., {Manso Sainz}, R., {Asensio Ramos}, A., \& {Belluzzi}, L. 2012, \href{http://dx.doi.org/10.1111/j.1365-2966.2011.19681.x}{\color{blue}\mnras}, \href{https://ui.adsabs.harvard.edu/abs/2012MNRAS.419..153M}{419, 153}

\bibitem[{{Mishra-Sharma} \& {Yang}(2022)}]{Mishra-Sharma2022mla}
{Mishra-Sharma}, S. \& {Yang}, G. 2022, in Machine Learning for Astrophysics, \href{https://ui.adsabs.harvard.edu/abs/2022mla..confE..34M}{34}

\bibitem[{{Morosin} {et~al.}(2022){Morosin}, {de la Cruz Rodr{\'\i}guez}, {D{\'\i}az Baso}, \& {Leenaarts}}]{Morosin2022A&A...664A...8M}
{Morosin}, R., {de la Cruz Rodr{\'\i}guez}, J., {D{\'\i}az Baso}, C.~J., \& {Leenaarts}, J. 2022, \href{http://dx.doi.org/10.1051/0004-6361/202243461}{\color{blue}\aap}, \href{https://ui.adsabs.harvard.edu/abs/2022A&A...664A...8M}{664, A8}

\bibitem[{{Morosin} {et~al.}(2020){Morosin}, {de la Cruz Rodr{\'\i}guez}, {Vissers}, \& {Yadav}}]{Morosin2020A&A}
{Morosin}, R., {de la Cruz Rodr{\'\i}guez}, J., {Vissers}, G. J.~M., \& {Yadav}, R. 2020, \href{http://dx.doi.org/10.1051/0004-6361/202038754}{\color{blue}\aap}, \href{https://ui.adsabs.harvard.edu/abs/2020A&A...642A.210M}{642, A210}

\bibitem[{Parikh \& Boyd(2014)}]{parikh_boyd14}
Parikh, N. \& Boyd, S. 2014, Foundations and Trends in Optimization, 1, 1

\bibitem[{{Pastor Yabar} {et~al.}(2021){Pastor Yabar}, {Borrero}, {Quintero Noda}, \& {Ruiz Cobo}}]{2021A&A...656L..20P}
{Pastor Yabar}, A., {Borrero}, J.~M., {Quintero Noda}, C., \& {Ruiz Cobo}, B. 2021, \href{http://dx.doi.org/10.1051/0004-6361/202142149}{\color{blue}\aap}, \href{https://ui.adsabs.harvard.edu/abs/2021A&A...656L..20P}{656, L20}

\bibitem[{{Paszke} {et~al.}(2019){Paszke}, {Gross}, {Massa}, {Lerer}, {Bradbury}, {Chanan}, {Killeen}, {Lin}, {Gimelshein}, {Antiga}, {Desmaison}, {K{\"o}pf}, {Yang}, {DeVito}, {Raison}, {Tejani}, {Chilamkurthy}, {Steiner}, {Fang}, {Bai}, \& {Chintala}}]{Paszke2019arXiv}
{Paszke}, A., {Gross}, S., {Massa}, F., {et~al.} 2019, \href{https://ui.adsabs.harvard.edu/abs/2019arXiv191201703P}{\href{http://dx.doi.org/10.48550/arXiv.1912.01703}{\color{blue}arXiv e-prints}, arXiv:1912.01703}

\bibitem[{{Pietrow} {et~al.}(2020){Pietrow}, {Kiselman}, {de la Cruz Rodr{\'\i}guez}, {D{\'\i}az Baso}, {Pastor Yabar}, \& {Yadav}}]{Pietrow2020A&A}
{Pietrow}, A.~G.~M., {Kiselman}, D., {de la Cruz Rodr{\'\i}guez}, J., {et~al.} 2020, \href{http://dx.doi.org/10.1051/0004-6361/202038750}{\color{blue}\aap}, \href{https://ui.adsabs.harvard.edu/abs/2020A&A...644A..43P}{644, A43}

\bibitem[{{Quintero Noda} {et~al.}(2022){Quintero Noda}, {Schlichenmaier}, {Bellot Rubio}, {L{\"o}fdahl}, {Khomenko}, {Jur{\v{c}}{\'a}k}, {Leenaarts}, {Kuckein}, {Gonz{\'a}lez Manrique}, {Gun{\'a}r}, {Nelson}, {de la Cruz Rodr{\'\i}guez}, {Tziotziou}, {Tsiropoula}, {Aulanier}, {Aboudarham}, {Allegri}, {Alsina Ballester}, {Amans}, {Asensio Ramos}, {Bail{\'e}n}, {Balaguer}, {Baldini}, {Balthasar}, {Barata}, {Barczynski}, {Barreto Cabrera}, {Baur}, {B{\'e}chet}, {Beck}, {Bel{\'\i}o-As{\'\i}n}, {Bello-Gonz{\'a}lez}, {Belluzzi}, {Bentley}, {Berdyugina}, {Berghmans}, {Berlicki}, {Berrilli}, {Berkefeld}, {Bettonvil}, {Bianda}, {Bienes P{\'e}rez}, {Bonaque-Gonz{\'a}lez}, {Braj{\v{s}}a}, {Bommier}, {Bourdin}, {Burgos Mart{\'\i}n}, {Calchetti}, {Calcines}, {Calvo Tovar}, {Campbell}, {Carballo-Mart{\'\i}n}, {Carbone}, {Carlin}, {Carlsson}, {Castro L{\'o}pez}, {Cavaller}, {Cavallini}, {Cauzzi}, {Cecconi}, {Chulani}, {Cirami}, {Consolini}, {Coretti}, {Cosentino}, {C{\'o}zar-Castellano}, {Dalmasse}, {Danilovic}, {De Juan
  Ovelar}, {Del Moro}, {del Pino Alem{\'a}n}, {del Toro Iniesta}, {Denker}, {Dhara}, {Di Marcantonio}, {D{\'\i}az Baso}, {Diercke}, {Dineva}, {D{\'\i}az-Garc{\'\i}a}, {Doerr}, {Doyle}, {Erdelyi}, {Ermolli}, {Escobar Rodr{\'\i}guez}, {Esteban Pozuelo}, {Faurobert}, {Felipe}, {Feller}, {Feijoo Amoedo}, {Femen{\'\i}a Castell{\'a}}, {Fernandes}, {Ferro Rodr{\'\i}guez}, {Figueroa}, {Fletcher}, {Franco Ordovas}, {Gafeira}, {Gardenghi}, {Gelly}, {Giorgi}, {Gisler}, {Giovannelli}, {Gonz{\'a}lez}, {Gonz{\'a}lez}, {Gonz{\'a}lez-Cava}, {Gonz{\'a}lez Garc{\'\i}a}, {G{\"o}m{\"o}ry}, {Gracia}, {Grauf}, {Greco}, {Grivel}, {Guerreiro}, {Guglielmino}, {Hammerschlag}, {Hanslmeier}, {Hansteen}, {Heinzel}, {Hern{\'a}ndez-Delgado}, {Hern{\'a}ndez Su{\'a}rez}, {Hidalgo}, {Hill}, {Hizberger}, {Hofmeister}, {J{\"a}gers}, {Janett}, {Jarolim}, {Jess}, {Jim{\'e}nez Mej{\'\i}as}, {Jolissaint}, {Kamlah}, {Kapit{\'a}n}, {Ka{\v{s}}parov{\'a}}, {Keller}, {Kentischer}, {Kiselman}, {Kleint}, {Klvana}, {Kontogiannis}, {Krishnappa},
  {Ku{\v{c}}era}, {Labrosse}, {Lagg}, {Landi Degl'Innocenti}, {Langlois}, {Lafon}, {Laforgue}, {Le Men}, {Lepori}, {Lepreti}, {Lindberg}, {Lilje}, {L{\'o}pez Ariste}, {L{\'o}pez Fern{\'a}ndez}, {L{\'o}pez Jim{\'e}nez}, {L{\'o}pez L{\'o}pez}, {Manso Sainz}, {Marassi}, {Marco de la Rosa}, {Marino}, {Marrero}, {Mart{\'\i}n}, {Mart{\'\i}n G{\'a}lvez}, {Mart{\'\i}n Hernando}, {Masciadri}, {Mart{\'\i}nez Gonz{\'a}lez}, {Matta-G{\'o}mez}, {Mato}, {Mathioudakis}, {Matthews}, {Mein}, {Merlos Garc{\'\i}a}, {Moity}, {Montilla}, {Molinaro}, {Molodij}, {Montoya}, {Munari}, {Murabito}, {N{\'u}{\~n}ez Cagigal}, {Oliviero}, {Orozco Su{\'a}rez}, {Ortiz}, {Padilla-Hern{\'a}ndez}, {Pa{\'e}z Ma{\~n}{\'a}}, {Paletou}, {Pancorbo}, {Pastor Ca{\~n}edo}, {Pastor Yabar}, {Peat}, {Pedichini}, {Peixinho}, {Pe{\~n}ate}, {P{\'e}rez de Taoro}, {Peter}, {Petrovay}, {Piazzesi}, {Pietropaolo}, {Pleier}, {Poedts}, {P{\"o}tzi}, {Podladchikova}, {Prieto}, {Quintero Nehrkorn}, {Ramelli}, {Ramos Sapena}, {Rasilla}, {Reardon}, {Rebolo}, {Regalado
  Olivares}, {Reyes Garc{\'\i}a-Talavera}, {Riethm{\"u}ller}, {Rimmele}, {Rodr{\'\i}guez Delgado}, {Rodr{\'\i}guez Gonz{\'a}lez}, {Rodr{\'\i}guez-Losada}, {Rodr{\'\i}guez Ramos}, {Romano}, {Roth}, {Rouppe van der Voort}, {Rudawy}, {Ruiz de Galarreta}, {Ryb{\'a}k}, {Salvade}, {S{\'a}nchez-Capuchino}, {S{\'a}nchez Rodr{\'\i}guez}, {Sangiorgi}, {Say{\`e}de}, {Scharmer}, {Scheiffelen}, {Schmidt}, {Schmieder}, {Scir{\`e}}, {Scuderi}, {Siegel}, {Sigwarth}, {Sim{\~o}es}, {Snik}, {Sliepen}, {Sobotka}, {Socas-Navarro}, {Sola La Serna}, {Solanki}, {Soler Trujillo}, {Soltau}, {Sordini}, {Sosa M{\'e}ndez}, {Stangalini}, {Steiner}, {Stenflo}, {{\v{S}}t{\v{e}}p{\'a}n}, {Strassmeier}, {Sudar}, {Suematsu}, {S{\"u}tterlin}, {Tallon}, {Temmer}, {Tenegi}, {Tritschler}, {Trujillo Bueno}, {Turchi}, {Utz}, {van Harten}, {van Noort}, {van Werkhoven}, {Vansintjan}, {Vaz Cedillo}, {Vega Reyes}, {Verma}, {Veronig}, {Viavattene}, {Vitas}, {V{\"o}gler}, {von der L{\"u}he}, {Volkmer}, {Waldmann}, {Walton}, {Wisniewska}, {Zeman},
  {Zeuner}, {Zhang}, {Zuccarello}, \& {Collados}}]{2022A&A...666A..21Q}
{Quintero Noda}, C., {Schlichenmaier}, R., {Bellot Rubio}, L.~R., {et~al.} 2022, \href{http://dx.doi.org/10.1051/0004-6361/202243867}{\color{blue}\aap}, \href{https://ui.adsabs.harvard.edu/abs/2022A&A...666A..21Q}{666, A21}

\bibitem[{{Rahaman} {et~al.}(2018){Rahaman}, {Baratin}, {Arpit}, {Draxler}, {Lin}, {Hamprecht}, {Bengio}, \& {Courville}}]{Rahaman2018arXiv}
{Rahaman}, N., {Baratin}, A., {Arpit}, D., {et~al.} 2018, \href{https://ui.adsabs.harvard.edu/abs/2018arXiv180608734R}{\href{http://dx.doi.org/10.48550/arXiv.1806.08734}{\color{blue}arXiv e-prints}, arXiv:1806.08734}

\bibitem[{{Rimmele} {et~al.}(2020){Rimmele}, {Warner}, {Keil}, {Goode}, {Kn{\"o}lker}, {Kuhn}, {Rosner}, {McMullin}, {Casini}, {Lin}, {W{\"o}ger}, {von der L{\"u}he}, {Tritschler}, {Davey}, {de Wijn}, {Elmore}, {Fehlmann}, {Harrington}, {Jaeggli}, {Rast}, {Schad}, {Schmidt}, {Mathioudakis}, {Mickey}, {Anan}, {Beck}, {Marshall}, {Jeffers}, {Oschmann}, {Beard}, {Berst}, {Cowan}, {Craig}, {Cross}, {Cummings}, {Donnelly}, {de Vanssay}, {Eigenbrot}, {Ferayorni}, {Foster}, {Galapon}, {Gedrites}, {Gonzales}, {Goodrich}, {Gregory}, {Guzman}, {Guzzo}, {Hegwer}, {Hubbard}, {Hubbard}, {Johansson}, {Johnson}, {Liang}, {Liang}, {McQuillen}, {Mayer}, {Newman}, {Onodera}, {Phelps}, {Puentes}, {Richards}, {Rimmele}, {Sekulic}, {Shimko}, {Simison}, {Smith}, {Starman}, {Sueoka}, {Summers}, {Szabo}, {Szabo}, {Wampler}, {Williams}, \& {White}}]{Rimmele2020SoPh_dkist}
{Rimmele}, T.~R., {Warner}, M., {Keil}, S.~L., {et~al.} 2020, \href{http://dx.doi.org/10.1007/s11207-020-01736-7}{\color{blue}\solphys}, \href{https://ui.adsabs.harvard.edu/abs/2020SoPh..295..172R}{295, 172}

\bibitem[{{Rouppe van der Voort} {et~al.}(2023){Rouppe van der Voort}, {van Noort}, \& {de la Cruz Rodr{\'\i}guez}}]{Rouppe2023A&A...673A..11R}
{Rouppe van der Voort}, L. H.~M., {van Noort}, M., \& {de la Cruz Rodr{\'\i}guez}, J. 2023, \href{http://dx.doi.org/10.1051/0004-6361/202345933}{\color{blue}\aap}, \href{https://ui.adsabs.harvard.edu/abs/2023A&A...673A..11R}{673, A11}

\bibitem[{{Saragadam} {et~al.}(2022){Saragadam}, {Tan}, {Balakrishnan}, {Baraniuk}, \& {Veeraraghavan}}]{Saragadam2022arXiv220203532S}
{Saragadam}, V., {Tan}, J., {Balakrishnan}, G., {Baraniuk}, R.~G., \& {Veeraraghavan}, A. 2022, \href{https://ui.adsabs.harvard.edu/abs/2022arXiv220203532S}{\href{http://dx.doi.org/10.48550/arXiv.2202.03532}{\color{blue}arXiv e-prints}, arXiv:2202.03532}

\bibitem[{{Scharmer} {et~al.}(2003){Scharmer}, {Bjelksjo}, {Korhonen}, {Lindberg}, \& {Petterson}}]{Scharmer2003}
{Scharmer}, G.~B., {Bjelksjo}, K., {Korhonen}, T.~K., {Lindberg}, B., \& {Petterson}, B. 2003, in \procspie, Vol. 4853, Innovative Telescopes and Instrumentation for Solar Astrophysics, ed. S.~L. {Keil} \& S.~V. {Avakyan}, \href{http://adsabs.harvard.edu/abs/2003SPIE.4853..341S}{341--350}

\bibitem[{{Scharmer} {et~al.}(2008){Scharmer}, {Narayan}, {Hillberg}, {de la Cruz Rodriguez}, {L{\"o}fdahl}, {Kiselman}, {S{\"u}tterlin}, {van Noort}, \& {Lagg}}]{Scharmer2008}
{Scharmer}, G.~B., {Narayan}, G., {Hillberg}, T., {et~al.} 2008, \href{http://dx.doi.org/10.1086/595744}{\color{blue}\apjl}, \href{http://adsabs.harvard.edu/abs/2008ApJ...689L..69S}{689, L69}

\bibitem[{{Sitzmann} {et~al.}(2020){Sitzmann}, {Martel}, {Bergman}, {Lindell}, \& {Wetzstein}}]{Sitzmann2020arXiv}
{Sitzmann}, V., {Martel}, J. N.~P., {Bergman}, A.~W., {Lindell}, D.~B., \& {Wetzstein}, G. 2020, \href{https://ui.adsabs.harvard.edu/abs/2020arXiv200609661S}{\href{http://dx.doi.org/10.48550/arXiv.2006.09661}{\color{blue}arXiv e-prints}, arXiv:2006.09661}

\bibitem[{{Sukhorukov} \& {Leenaarts}(2017)}]{2017A&A...597A..46S}
{Sukhorukov}, A.~V. \& {Leenaarts}, J. 2017, \href{http://dx.doi.org/10.1051/0004-6361/201629086}{\color{blue}\aap}, \href{https://ui.adsabs.harvard.edu/abs/2017A&A...597A..46S}{597, A46}

\bibitem[{{Tancik} {et~al.}(2020){Tancik}, {Srinivasan}, {Mildenhall}, {Fridovich-Keil}, {Raghavan}, {Singhal}, {Ramamoorthi}, {Barron}, \& {Ng}}]{Tancik2020arXiv}
{Tancik}, M., {Srinivasan}, P.~P., {Mildenhall}, B., {et~al.} 2020, \href{https://ui.adsabs.harvard.edu/abs/2020arXiv200610739T}{\href{http://dx.doi.org/10.48550/arXiv.2006.10739}{\color{blue}arXiv e-prints}, arXiv:2006.10739}

\bibitem[{{van Noort}(2012)}]{vanNoort2012A&A}
{van Noort}, M. 2012, \href{http://dx.doi.org/10.1051/0004-6361/201220220}{\color{blue}\aap}, \href{https://ui.adsabs.harvard.edu/abs/2012A&A...548A...5V}{548, A5}

\bibitem[{{van Noort} {et~al.}(2022){van Noort}, {Bischoff}, {Kramer}, {Solanki}, \& {Kiselman}}]{vanNoort2022A&A...668A.149V}
{van Noort}, M., {Bischoff}, J., {Kramer}, A., {Solanki}, S.~K., \& {Kiselman}, D. 2022, \href{http://dx.doi.org/10.1051/0004-6361/202243464}{\color{blue}\aap}, \href{https://ui.adsabs.harvard.edu/abs/2022A&A...668A.149V}{668, A149}

\bibitem[{{Vissers} {et~al.}(2021){Vissers}, {Danilovic}, {de la Cruz Rodr{\'\i}guez}, {Leenaarts}, {Morosin}, {D{\'\i}az Baso}, {Reid}, {Pomoell}, {Price}, \& {Inoue}}]{Vissers2021A&A}
{Vissers}, G.~J.~M., {Danilovic}, S., {de la Cruz Rodr{\'\i}guez}, J., {et~al.} 2021, \href{http://dx.doi.org/10.1051/0004-6361/202038900}{\color{blue}\aap}, \href{https://ui.adsabs.harvard.edu/abs/2021A&A...645A...1V}{645, A1}

\bibitem[{{Vissers} {et~al.}(2022){Vissers}, {Danilovic}, {Zhu}, {Leenaarts}, {D{\'\i}az Baso}, {da Silva Santos}, {de la Cruz Rodr{\'\i}guez}, \& {Wiegelmann}}]{Vissers2022A&A_wfavsmhm}
{Vissers}, G.~J.~M., {Danilovic}, S., {Zhu}, X., {et~al.} 2022, \href{http://dx.doi.org/10.1051/0004-6361/202142087}{\color{blue}\aap}, \href{https://ui.adsabs.harvard.edu/abs/2022A&A...662A..88V}{662, A88}

\bibitem[{{{\v{S}}t{\v{e}}p{\'a}n} {et~al.}(2022){{\v{S}}t{\v{e}}p{\'a}n}, {del Pino Alem{\'a}n}, \& {Trujillo Bueno}}]{Jiri2022A&A...659A.137S}
{{\v{S}}t{\v{e}}p{\'a}n}, J., {del Pino Alem{\'a}n}, T., \& {Trujillo Bueno}, J. 2022, \href{http://dx.doi.org/10.1051/0004-6361/202142079}{\color{blue}\aap}, \href{https://ui.adsabs.harvard.edu/abs/2022A&A...659A.137S}{659, A137}

\bibitem[{{{\v{S}}t{\v{e}}p{\'a}n} {et~al.}(2024){{\v{S}}t{\v{e}}p{\'a}n}, {del Pino Alem{\'a}n}, \& {Trujillo Bueno}}]{Jiri2024arXiv240720926S}
{{\v{S}}t{\v{e}}p{\'a}n}, J., {del Pino Alem{\'a}n}, T., \& {Trujillo Bueno}, J. 2024, \href{https://ui.adsabs.harvard.edu/abs/2024arXiv240720926S}{\href{http://dx.doi.org/10.48550/arXiv.2407.20926}{\color{blue}arXiv e-prints}, arXiv:2407.20926}

\bibitem[{{Wiegelmann} {et~al.}(2008){Wiegelmann}, {Thalmann}, {Schrijver}, {De Rosa}, \& {Metcalf}}]{Wiegelmann2008SoPh..247..249W}
{Wiegelmann}, T., {Thalmann}, J.~K., {Schrijver}, C.~J., {De Rosa}, M.~L., \& {Metcalf}, T.~R. 2008, \href{http://dx.doi.org/10.1007/s11207-008-9130-y}{\color{blue}\solphys}, \href{https://ui.adsabs.harvard.edu/abs/2008SoPh..247..249W}{247, 249}

\bibitem[{{Yadav} {et~al.}(2021){Yadav}, {D{\'\i}az Baso}, {de la Cruz Rodr{\'\i}guez}, {Calvo}, \& {Morosin}}]{Yadav2021A&A...649A.106Y}
{Yadav}, R., {D{\'\i}az Baso}, C.~J., {de la Cruz Rodr{\'\i}guez}, J., {Calvo}, F., \& {Morosin}, R. 2021, \href{http://dx.doi.org/10.1051/0004-6361/202039857}{\color{blue}\aap}, \href{https://ui.adsabs.harvard.edu/abs/2021A&A...649A.106Y}{649, A106}

\bibitem[{{Yadav} {et~al.}(2023){Yadav}, {Kazachenko}, {Afanasyev}, {de la Cruz Rodr{\'\i}guez}, \& {Leenaarts}}]{Yadav2023ApJ_ERF}
{Yadav}, R., {Kazachenko}, M.~D., {Afanasyev}, A.~N., {de la Cruz Rodr{\'\i}guez}, J., \& {Leenaarts}, J. 2023, \href{http://dx.doi.org/10.3847/1538-4357/acfd2b}{\color{blue}\apj}, \href{https://ui.adsabs.harvard.edu/abs/2023ApJ...958...54Y}{958, 54}

\end{thebibliography}

\appendix
\section{Additional figures}

\begin{figure*}[htp!]
\centering
\includegraphics[width=1\linewidth]{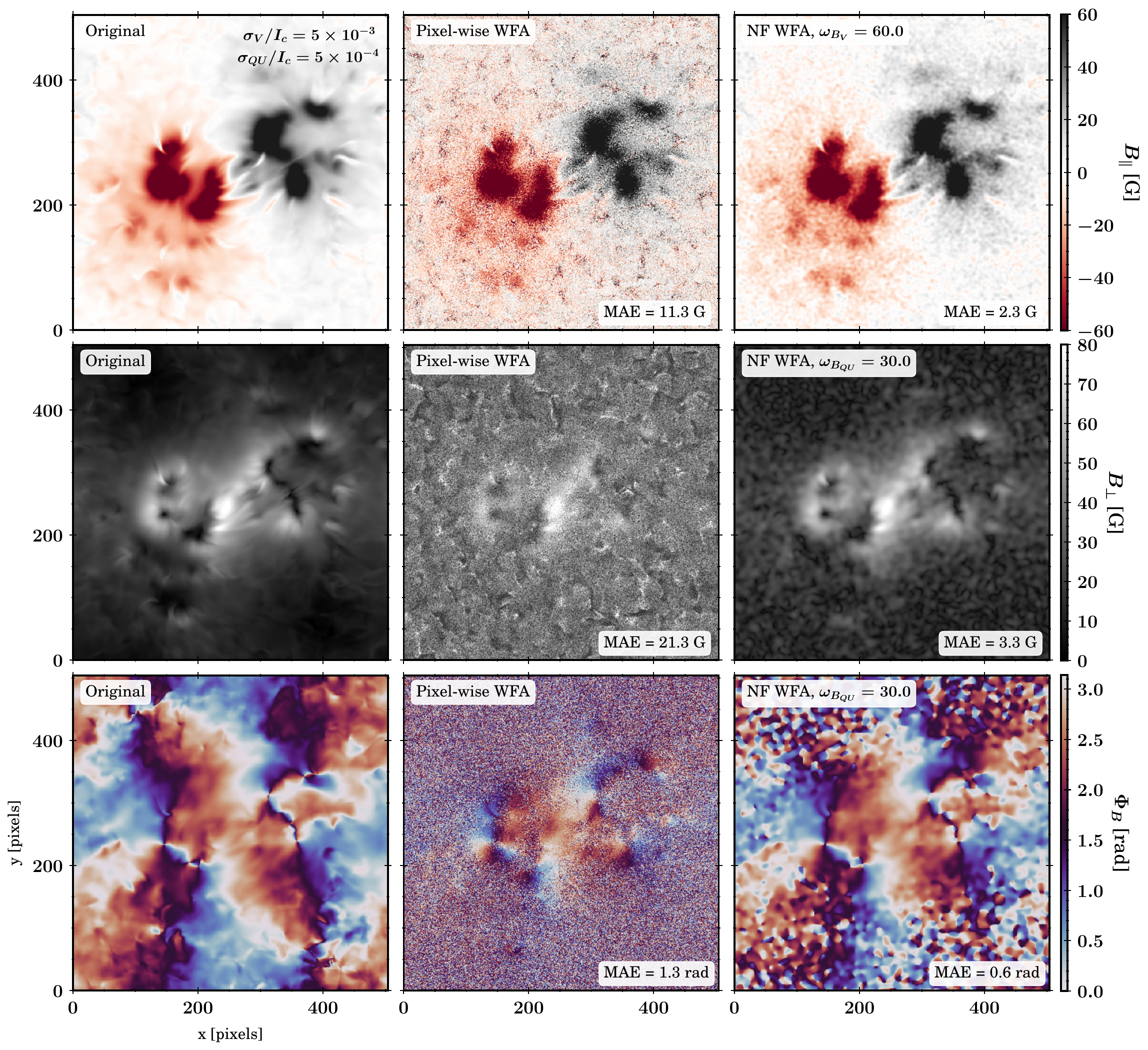}
\caption{Same as Fig.~\ref{fig:simu_mosaic} for an increased noise level. From left to right: original magnetic field from the simulation, magnetic field inferred by the pixel-wise WFA and results using the WFA NF.} \label{fig:simu_mosaic_appendix}
\end{figure*}


\begin{figure}[htp!]
\centering
\includegraphics[width=\linewidth]{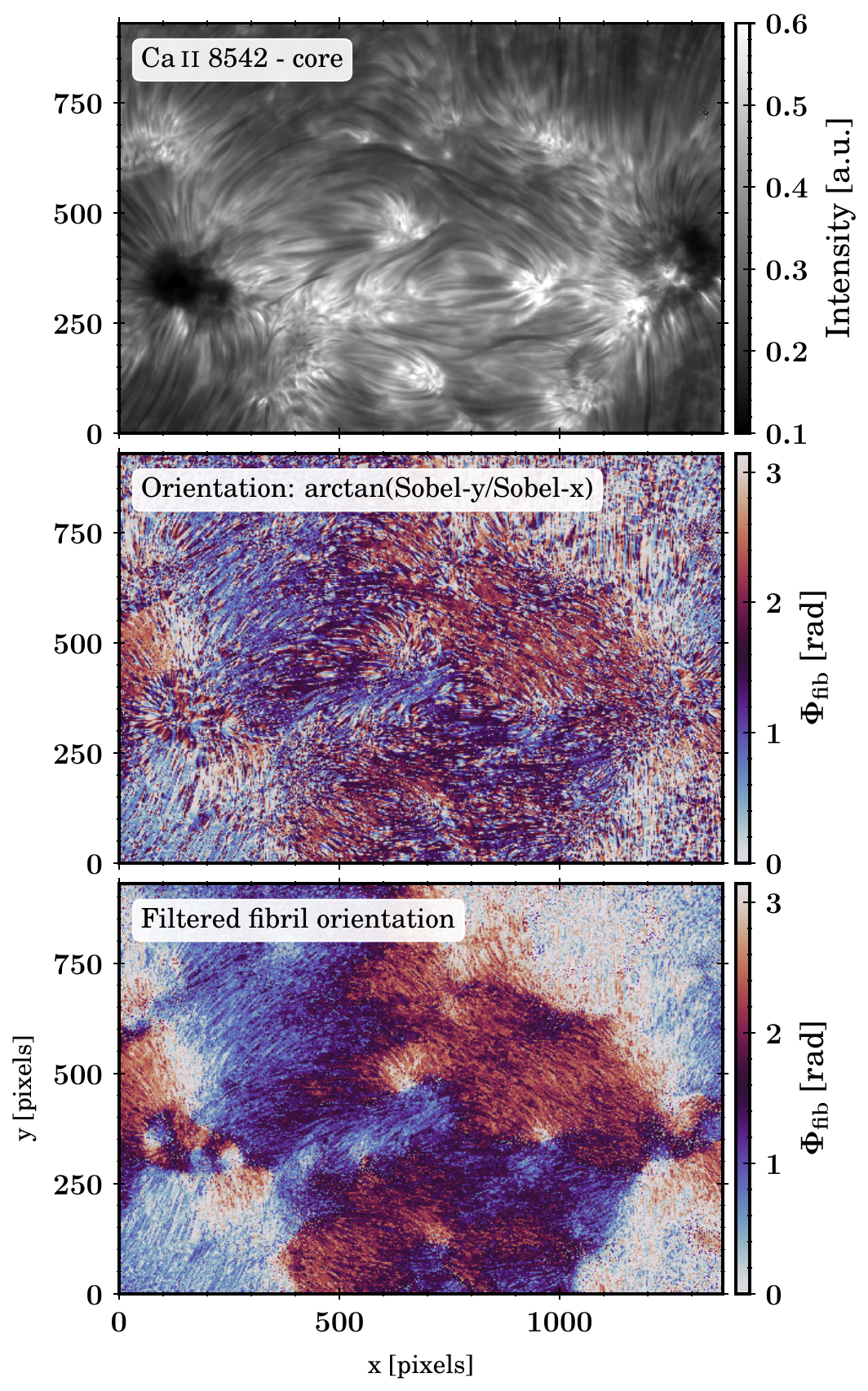}
\caption{Steps to retrieve the orientation of the fibrils using the intensity of the \ion{Ca}{ii} line: the orientation (middle panel) is obtained by applying Sobel filters to the monochromatic image at the core (top panel), which is later filtered (bottom panel).} \label{fig:orientation}
\end{figure}

\end{document}